\documentclass{LMCS}

\def\doi{8(4:2)2012}
\lmcsheading%
{\doi}
{1--27}
{}
{}
{Jan.~31, 2012}
{Oct.~\phantom05, 2012}
{}

\usepackage{enumerate}

\usepackage{amsmath,amssymb,amsthm} % AMS-LaTeX
\usepackage{prooftree,diagrams}

\usepackage{hyperref}

\newcommand{\comptype}[1]{\underline{#1}}

\newcommand{\VconstA}{\alpha}
\newcommand{\VconstB}{\beta}

\newcommand{\CconstA}{\comptype{\alpha}}
\newcommand{\CconstB}{\comptype{\beta}}

\newcommand{\VA}{\mathsf{A}}
\newcommand{\VB}{\mathsf{B}}
\newcommand{\VC}{\mathsf{C}}

\newcommand{\CA}{\comptype{\mathsf{A}}}
\newcommand{\CB}{\comptype{\mathsf{B}}}
\newcommand{\CC}{\comptype{\mathsf{C}}}
\newcommand{\CD}{\comptype{\mathsf{D}}}

\newcommand{\CR}{\comptype{\mathsf{R}}}
\newcommand{\CI}{\comptype{\mathsf{I}}}

\newcommand{\Vone}{1}
\newcommand{\Vprod}{\times}
\newcommand{\Vfun}{\to}

\newcommand{\lpop}{\multimap}

\newcommand{\Cone}{\comptype{1}}
\newcommand{\Cprod}{\,\&\,}
\newcommand{\Cfun}{\Rightarrow}
\newcommand{\Cbang}[1]{{! #1}}  
\newcommand{\Ccopower}[2]{! #1 \, {\otimes} \, #2}
\newcommand{\Czero}{\comptype{0}}
\newcommand{\Cplus}{\oplus}

\newcommand{\In}[2]{#1 \colon  \! #2}
\newcommand{\rIn}[2]{#1 \colon  #2}

\newcommand{\Cj}[4]{#1 \mid  \! #2 \, \vdash \, \rIn{#3}{#4}}
\newcommand{\Vj}[3]{\Cj{#1}{{-}}{#2}{#3}}

\newcommand{\Ceq}[5]{#1 \mid  \! #2 \, \vdash \, #3 = #4 \colon #5}
\newcommand{\Veq}[4]{\Ceq{#1}{-}{#2}{#3}{#4}}

\newcommand{\LCequals}{=_{\lambda_c}}
\newcommand{\BEequals}{=_{\beta\eta}}

\newcommand{\LCeq}[4]{#1 \, \vdash \, #2 =_{\lambda_c} #3 \colon #4}
\newcommand{\BEeq}[4]{#1 \, \vdash \, #2 =_{\beta\eta} #3 \colon #4}

\newcommand{\Csim}[5]{#1 \mid  \! #2 \, \vdash \, #3 \sim #4 \colon #5}
\newcommand{\Vsim}[4]{\Csim{#1}{-}{#2}{#3}{#4}}

\newcommand{\Vstar}{{*}}
\newcommand{\Vpair}[2]{\langle #1 , #2 \rangle}
\newcommand{\Vfst}[1]{\mathrm{fst}(#1)}
\newcommand{\Vsnd}[1]{\mathrm{snd}(#1)}

\newcommand{\Vlam}[3]{\lambda \In{#1}{#2}.\: #3}
\newcommand{\Vappl}[2]{#1(#2)}

\newcommand{\compop}[1]{\underline{#1}}

\newcommand{\Cstar}{\compop{*}}
\newcommand{\Cpair}[2]{\compop{\langle} #1 , #2 \compop{\rangle}}
\newcommand{\Cfst}[1]{\compop{\mathrm{fst}}(#1)}
\newcommand{\Csnd}[1]{\compop{\mathrm{snd}}(#1)}

\newcommand{\Clam}[3]{\compop{\lambda} \In{#1}{#2}.\: #3}
\newcommand{\Cappl}[2]{#1\compop{(}#2\compop{)}}

\newcommand{\bang}[1]{{! #1}}  
\newcommand{\Itop}{\top}
\newcommand{\Ilet}[2]{\mathrm{let}\: {\Itop}\:\mathrm{be}\:{#1} \;\mathrm{in}\: #2}
\newcommand{\banglet}[3]{\mathrm{let}\: {\bang #1}\:\mathrm{be}\:{#2} \;\mathrm{in}\: #3}

\newcommand{\llambda}{\lambda^{\!\circ\!}}
\newcommand{\llam}[3]{\llambda \In{#1}{#2}.\: #3}
\newcommand{\lappl}[2]{#1[#2]}

\newcommand{\copowerterm}[2]{{\bang{#1}}\! \otimes \! #2}
\newcommand{\copowerlet}[4]{\mathrm{let}\: {\copowerterm{#1}{#2}}\:\mathrm{be}\:{#3} \;\mathrm{in}\: #4}

\newcommand{\Cimage}[1]{\compop{?}(#1)}
\newcommand{\Cimagetype}[2]{\compop{?}_{#1}(#2)}
\newcommand{\Cinl}[1]{\compop{\mathrm{inl}}(#1)}
\newcommand{\Cinr}[1]{\compop{\mathrm{inr}}(#1)}
\newcommand{\Ccase}[5]{\compop{\mathrm{case}} \, #1 \,\mathrm{of}\,( \Cinl{#2}. \, #3 ; \, \Cinr{#4}. \, #5)}

\newcommand{\Lone}{1}
\newcommand{\Lprod}{\times}
\newcommand{\Lfun}{\to}

\newcommand{\Lj}[3]{#1  \, \vdash  \rIn{#2}{#3}}

\newcommand{\Lstar}{*}
\newcommand{\Lpair}[2]{\langle #1 , #2 \rangle}
\newcommand{\Lfst}[1]{\mathrm{fst}(#1)}
\newcommand{\Lsnd}[1]{\mathrm{snd}(#1)}

\newcommand{\Llam}[3]{\lambda \In{#1}{#2}.\: #3}
\newcommand{\Lappl}[2]{#1  #2}

\newcommand{\cbv}[1]{#1^{\mathrm{v}}}
\newcommand{\cbn}[1]{#1^{\mathrm{n}}}

\newcommand{\cbvLincps}[1]{#1^{\mathrm{v_{\CR}}}}
\newcommand{\cbnLincps}[1]{#1^{\mathrm{n_{\CR}}}}

\newcommand{\CpsVT}[1]{#1^{\mathcal{V}_{\CR}}}
\newcommand{\CpsCT}[1]{#1^{\mathcal{C}_{\CR}}}

\newcommand{\CpsVTcbv}[1]{\CpsVT{(\cbv{#1})}}
\newcommand{\CpsCTcbn}[1]{\CpsCT{(\cbn{#1})}}
\newcommand{\CpsVTcbn}[1]{\CpsVT{(\cbn{#1})}}

\newcommand{\CpsVVT}[1]{#1^{\mathcal{V}_{\CR}\mathcal{V}_{\CR}}}
\newcommand{\CpsCCT}[1]{#1^{\mathcal{C}_{\CR}\mathcal{C}_{\CR}}}
\newcommand{\CpsVVVT}[1]{#1^{\mathcal{V}_{\CR}\mathcal{V}_{\CR}\mathcal{V}_{\CR}}}

\newcommand{\Viso}[1]{i_{#1}}
\newcommand{\Ciso}[1]{j_{#1}}

\newcommand{\gspace}{2ex} % gather newline spacing

\begin{document}

%\frontmatter          % for the preliminaries
%
% \pagestyle{headings}  % switches on printing of running heads

\title[Linear-use CPS Translations in EEC]{Linear-use CPS Translations \\ in the Enriched Effect Calculus\rsuper*}

\author[J.~Egger]{Jeff Egger\rsuper a}  
\address{{\lsuper a}Department of Physics and Atmospheric Science,
Dalhousie University, Halifax, N.S., Canada}
\email{jeffegger@yahoo.ca}
\thanks{{\lsuper a}Research carried out while Egger was at 
LFCS, University of Edinburgh.
Research supported by EPSRC Research Grant ``Linear Observations and Computational Effects'',
and by the Danish Agency for Science, Technology and Innovation.}

\author[R.~E.~M{\o}gelberg]{Rasmus Ejlers M{\o}gelberg\rsuper b}
\address{{\lsuper b}IT University of Copenhagen, Copenhagen,  Denmark}
\email{mogel@itu.dk}

\author[A.~Simpson]{Alex Simpson\rsuper c}
\address{{\lsuper c}LFCS, School of Informatics, University of Edinburgh, Scotland, UK}
\email{Alex.Simpson@ed.ac.uk}



\titlecomment{{\lsuper*}The results in this paper first appeared 
in the proceedings of FoSSaCS 2010, \cite{EMS:fossacs}.}

\maketitle

\begin{abstract}
\noindent
The \emph{enriched effect calculus} (EEC) is an extension of Moggi's
computational metalanguage with a selection of primitives
from linear logic. This paper explores the enriched effect calculus
as a target language for continuation-passing-style (CPS) translations
in which the typing of the translations enforces the linear usage
of continuations. We first observe that 
established call-by-value and call-by name linear-use
CPS translations of simply-typed lambda-calculus into intuitionistic linear logic (ILL) 
land in the fragment of ILL given by EEC.
These two translations are
uniformly generalised by a single generic
translation of the enriched effect calculus into itself.
As our main theorem, we prove that the generic
self-translation of EEC is involutive up to isomorphism.
As corollaries, we obtain full completeness results,
both for the generic translation, and for 
the original call-by-value and call-by-name translations.
%The main syntactic theorem is proved using a category-theoretic
%semantics for the enriched effect calculus. We show that models are
%closed under a natural \emph{dual model} construction. The canonical
%linearly-used CPS translation then arises as
%the unique (up to isomorphism) map from the
%syntactic initial model to its own dual. This map
%is an equivalence of models.
%Thus the initial model is self-dual.
\end{abstract}

\maketitle

\section{Introduction}

Under a continuation-passing-style (CPS) interpretation, 
a call-by-value program
from $X$ to $Y$ is interpreted as a ``continuation transformer'',
that is, as a map $(Y \!\to\! R) \to (X \!\to \!R)$,  where $R$ represents the 
possible ``results'' of a computation.
Such maps are in one-to-one correspondence with Kleisi maps
for the \emph{continuations monad} $((-) \to R) \to R$, introduced by 
Moggi in~\cite{Moggi:89,Moggi:91}. 
In~\cite{BORT:02}, Berdine \emph{et al.}\ observe that,
in many programming situations, continuation transformers satisfy an
additional property: 
their argument, the continuation $Y \to R$, is used just once,
that is, it is used \emph{linearly}.
Thus a call-by-value program can be more informatively modelled as a
linear function $(Y \to R) \lpop (X \to R)$, corresponding to a
Kleisli map for the \emph{linearly-used continuations monad} 
$((-) \to R) \lpop R$.  

One goal of the present paper is to address the question: 
what is the natural type-theoretic context for
modelling linearly-used continuations? 
With the presence of both intuitionistic ($\to$) and linear ($\lpop$)
arrows, \emph{intuitionistic linear logic (ILL)}~\cite{Girard:87}
seems a natural answer.
Indeed, ILL has been used as the basis of a systematic study of 
linearly-used continuations by Hasegawa. 
%~\cite{Hasegawa:Flops:02,Hasegawa:Flops:04}.
In~\cite{Hasegawa:Flops:02}, he presents a
continuation passing style (CPS) translation of 
Moggi's call-by-value computational $\lambda$-calculus into
ILL, using the linearly-used continuations monad, and establishes
a full completeness result for this. 
A follow-up paper~\cite{Hasegawa:Flops:04} considers call-by-name.

In this paper we use a more general type theory, the \emph{enriched
  effect calculus (EEC)} introduced in~\cite{EMS,EMSb}, as a target
language for linear-use CPS translations.  On the one hand, EEC can be
seen as a fragment of ILL and, as such, its models strictly generalise
models of ILL.
% Thus
% our treatment of linearly-used continuations in EEC 
% broadens the realm of interpretations available for 
% linearly-used continuations.
On the other hand, it is a conservative extension of the standard
calculi for modelling computational effects (Moggi's
\emph{computational metalanguage}~\cite{Moggi:91}, 
and Levy's \emph{call-by-push-value (CBPV)}~\cite{Levy:book}) with a
selection of constructs from linear logic. 
In fact, any \emph{adjunction model} of CBPV~\cite{Levy:models}
(and hence any model of Moggi's computational metalanguage)
expands to a model of EEC~\cite{EMS,EMSc}. %, as shown in
%% Furthermore, many natural models of CBPV already possess the
%% structure needed to model EEC. 
% Thus the  treatment of linearly-used continuations in EEC means that
%any model of CBPV gives rise to an associated interpretation of
% the linearly-used continuations monad. 
%These facts provide
This provides an abundant supply of 
computationally interesting models of EEC
that are not models of ILL.

The paper 
begins with a brief presentation of the enriched effect
calculus, in Section~\ref{section:calculus}. 
The standard call-by-value and call-by-name translations of typed $\lambda$-calculus
into effect calculi (cf.~Moggi~\cite{Moggi:91},
Filinski~\cite{Filinski:phd}, Levy~\cite{Levy:book})
are then reviewed in Section~\ref{section:cbv:cbn},
using EEC as the target language.
This is followed, in Section~\ref{sec:lin:cps}, by giving corresponding
linear-use CPS translations within EEC.
The starting point is the observation that Hasegawa's
call-by-value~\cite{Hasegawa:Flops:02} 
and call-by-name~\cite{Hasegawa:Flops:04} 
linear-use CPS translations of simply-typed $\lambda$-calculus both
fall inside the fragment of ILL corresponding to EEC. 
One contribution of the paper is to show that, 
using EEC, we can recover these translations in a particularly
interesting way. 
This is achieved by identifying, in Section~\ref{section:canonical}, a single generic linear-use
CPS-translation of the entire enriched effect calculus into itself.
In Section~\ref{section:recovering}, it is shown how
Hasegawa's call-by-value and call-by-name translations are derived
from this by composing the generic translation with the standard  
call-by-value and call-by-name encodings of typed $\lambda$-calculus
into effect calculi, reviewed in Section~\ref{section:cbv:cbn}.

The generic linear-use CPS-translation of EEC into itself
is the principal contribution of the paper. It
possesses a remarkable property, unexpected in the context of
CPS translations: it is involutive up to isomorphism. 
That is, the translation of a translated term equals
the original term modulo type isomorphism. 
This property is stated as Theorem~\ref{theorem:involution}, 
which is the main theorem of the paper. 
As consequences, we obtain full-completeness results, both for
the generic self-translation itself (Theorem~\ref{thm:full:complete}),
and also for the
call-by-value and call-by-name linear-use CPS translations
into EEC, % Theorem~\ref{theorem:hassei:analogue},
mirroring Hasegawa's results for 
the translations into ILL. % ~\cite{Hasegawa:Flops:02}.

In the conference presentation of these results~\cite{EMS:fossacs},
the main syntactic theorem was given a semantic proof using  
category-theoretic models of EEC. In contrast, 
in the present paper, we provide purely syntactic proofs of all results.
It is hoped that this decision  will enlarge the potential readership of
the paper. Nevertheless, in Section~\ref{section:perspectives}, we briefly
outline the semantic context within which the syntactic results can be understood.
Even at an informal level, the semantic picture  provides an illuminating perspective
on the definition and properties of the generic self-translation of EEC.
A full treatment of the semantic side, which requires
considerable technical machinery, will be presented in a 
companion paper~\cite{EMSc}, devoted entirely to the category-theoretic model
theory of EEC.

A few words on the style of the paper. Since the presentation
is syntactic, there are many proofs by induction. Some of
these have numerous cases. (The proof of Theorem~\ref{theorem:involution}, for example,
has 41 cases.) In order to keep the paper concise and readable, in  such proofs, we 
present only a few illustrative cases, including the most interesting.
However, we take care to establish all the side results (for example, the
substitution property of Proposition~\ref{prop:trans:subs})  needed to make completing
the main proofs routine in principle (if lengthy in practice).

\section{The enriched effect calculus}
\label{section:calculus}

The \emph{enriched effect calculus (EEC)}~\cite{EMS,EMSb} is an extension of Moggi's 
computational metalanguage~\cite{Moggi:91}
with constructors from linear type theory. 
Similar to Filinski's effect PCF~\cite{Filinski:phd}
and Levy's CBPV~\cite{Levy:book}, it has two notions of types: \emph{value types} and
\emph{computation types}.
We use $\VconstA, \VconstB, \dots$ to range over a 
set of \emph{value type constants}, and
$\CconstA,\CconstB,\dots$ to range over a 
disjoint set of \emph{computation type constants}.
We then use $\VA,\VB,\dots$ to range over 
\emph{value types}, and 
$\CA,\CB,\dots$ to range over \emph{computation types},
which are specified by the grammar below.
\begin{align*}
\VA \, & ::= \, \VconstA \,\mid \,\Vone \,\mid \,\VA \Vprod \VB \,\mid \,\VA \Vfun \VB \,\mid \, \CA \, \mid \, \CA  \lpop \CB \\
\CA \, & ::= \, \CconstA \,\mid \,\Cone \,\mid \,\CA \Cprod \CB \,\mid\, \VA \Cfun \CB \,\mid \, \CI \,\mid \, \Cbang{\VA} \mid  \, \Ccopower{\VA}{\CB} \, \mid \, \Czero \, \mid \, \CA \Cplus \CB \enspace .
\end{align*}

As in~\cite{EMS,EMSb}, our notation has been heavily influenced by linear logic.
Indeed, EEC can be roughly understood as a fragment of intuitionistic linear logic.
However, there are some discrepancies, both in content and in syntax.
An important difference is that, in EEC, computation types are the sole source of
linearity. Thus linear function space $\CA \lpop \CB$ is defined between computation types only. 
However, the type $\CA \lpop \CB$ itself is a value type not a computation type.
As discused in \emph{op.\ cit.}, this choice seems
essential for EEC to be compatible with arbitrary (possibly non-commutative) computational effects.
A consequence is that the linear
function space cannot be iterated 
(neither $(\CA \lpop \CB) \lpop \CC$ nor $\CA \lpop (\CB \lpop \CC)$ is allowed).

Concerning notation, 
we remark that the type $\Ccopower{\VA}{\CB}$ is obtained by the application
of a single primitive binary type constructor $\Ccopower{(-)}{(-)}$ to a value type $\VA$ and 
computation type $\CB$. The hybrid notation for this constructor is chosen to emphasise the connection
with linear logic. 
In the present paper, we distinguish notationally
between products of computation types $\Cone$ and $\CA \Cprod \CB$, and products of value
types $\Vone$ and $\VA \Vprod \VB$. Similarly, we distinguish notationally between
computation-type function types $\VA \Cfun \CB$ 
(note that the the domain is a value type) and value-type function types
$\VA \Vfun \VB$. These choices, while adding redundancy to the streamlined syntax of~\cite{EMS,EMSb},
have the advantage of  simplifying certain properties of the syntactic 
translations we shall give in Section~\ref{sec:lin:cps}.
A further redundancy, introduced to simplify the presentation in Section~\ref{section:canonical},
is that we introduce a primitive computation type $\CI$, which plays a role analogous
to the tensor-product unit in linear logic.\footnote{Our choice of notation for units differs from
that of linear logic. In linear logic, the tensor unit, which we call $\CI$,  is written $1$,
and the  unit of the linear product $\&$, which we call $\Cone$, is written $\top$.}
This is redundant because $\CI$ can be defined as $\Cbang{\Vone}$.
As in linear logic, in addition to the linear isomorphism 
$\CI \cong {\Cbang{\Vone}}$, the type $\CI$ enjoys the further
isomorphisms $\Cbang{\VA} \cong {\Ccopower{\VA}{\CI}}$, and
$\CA \cong \CI \lpop \CA$ in EEC
(the latter isomorphism is not linear, since $\CI \lpop \CA$ is not a computation
type).
Finally, 
in EEC, the exponential type $\bang{\VA}$ plays the role of 
Moggi's monadic type $T \VA$ and Levy's type $F \VA$. The linear exponential
notation is motivated by the many formal analogies between the properties of $\bang{(-)}$
in EEC and in ILL. For example, EEC has the type isomorphisms
$\VA \Cfun \CB \cong \Cbang{\VA} \lpop \CB \cong  \VA \Vfun \CB$ (although only
the first is a computation type).
As in~\cite{EMS,EMSb},
we choose to make  
Levy's $U$ type constructor (see \cite{Levy:book}) invisible by including
computation types as value types. 
%The reader is referred to~\cite{EMS,EMSb} for further 
%discussion of enriched-effect-calculus types.

% and for more detailed comparisons with  other calculi.

\begin{figure}  % [t]
\vspace*{30pt}
\begin{gather*}
\prooftree
\justifies
\Vj{\Gamma,\, \In{x}{\VA}}{x}{\VA}
\endprooftree
\qquad
\prooftree
\justifies 
\Vj{\Gamma}{\Vstar}{\Vone}
\endprooftree
\\[\gspace]
\prooftree
\Vj{\Gamma}{t}{\VA}
  \quad
\Vj{\Gamma}{u}{\VB}
\justifies 
\Vj{\Gamma}{\Vpair{t}{u}}{\VA \Vprod \VB}
\endprooftree
\qquad
\prooftree
\Vj{\Gamma}{t}{\VA \Vprod \VB}
\justifies 
\Vj{\Gamma}{\Vfst{t}}{\VA}
\endprooftree
\qquad
\prooftree
\Vj{\Gamma}{t}{\VA \Vprod \VB}
\justifies 
\Vj{\Gamma}{\Vsnd{t}}{\VB}
\endprooftree
\\[\gspace]
\prooftree
\Vj{\Gamma,\,\In{x}{\VA}}{t}{\VB}
\justifies
\Vj{\Gamma}{\Vlam{x}{\VA}{t}}{\VA \Vfun \VB}
\endprooftree
\qquad
\prooftree
\Vj{\Gamma}{s}{\VA \Vfun \VB} 
  \quad
\Vj{\Gamma}{t}{\VA} 
\justifies
\Vj{\Gamma}{\Vappl{s}{t}}{\VB}
\endprooftree
\\[\gspace]
\prooftree
\justifies
\Cj{\Gamma}{\In{z}{\CA}}{z}{\CA}
\endprooftree
\qquad
\prooftree
\justifies 
\Cj{\Gamma}{\Delta}{\Cstar}{\Cone}
\endprooftree
\\[\gspace]
\prooftree
\Cj{\Gamma}{\Delta}{t}{\CA}
 \quad
\Cj{\Gamma}{\Delta}{u}{\CB}
\justifies 
\Cj{\Gamma}{\Delta}{\Cpair{t}{u}}{\CA \Cprod \CB}
\endprooftree
\qquad
\prooftree
\Cj{\Gamma}{\Delta}{t}{\CA \Cprod \CB}
\justifies 
\Cj{\Gamma}{\Delta}{\Cfst{t}}{\CA}
\endprooftree
\qquad
\prooftree
\Cj{\Gamma}{\Delta}{t}{\CA \Cprod \CB}
\justifies 
\Cj{\Gamma}{\Delta}{\Csnd{t}}{\CB}
\endprooftree
\\[\gspace]
\prooftree
\Cj{\Gamma,\,\In{x}{\VA}}{\Delta}{t}{\CB}
\justifies
\Cj{\Gamma}{\Delta}{\Clam{x}{\VA}{t}}{\VA \Cfun \CB}
\endprooftree
\qquad
\prooftree
\Cj{\Gamma}{\Delta}{s}{\VA \Cfun \CB} 
  \quad
\Vj{\Gamma}{t}{\VA} 
\justifies
\Cj{\Gamma}{\Delta}{\Cappl{s}{t}}{\CB}
\endprooftree
\\[\gspace]
\prooftree
\justifies
\Vj{\Gamma}{\Itop}{\CI}
\endprooftree
\qquad
\prooftree
\Cj{\Gamma}{\Delta}{t}{\CI}
\quad
\Vj{\Gamma}{u}{\CA}
\justifies
\Cj{\Gamma}{\Delta}{\Ilet{t}{u}}{\CA}
\endprooftree
\\[\gspace]
\prooftree
\Vj{\Gamma}{t}{\VA}
\justifies 
\Vj{\Gamma}{\bang{t}}{\Cbang{\VA}}
\endprooftree
\qquad
\prooftree
\Cj{\Gamma}{\Delta}{t}{\Cbang{\VA}}
\quad
\Vj{\Gamma,\, \In{x}{\VA}}{u}{\CB}
\justifies
\Cj{\Gamma}{\Delta}{\banglet{x}{t}{u}}{\CB}
\endprooftree
\\[\gspace]
\prooftree
\Vj{\Gamma}{t}{\VA}
\quad
\Cj{\Gamma}{\Delta}{u}{\CB}
\justifies
\Cj{\Gamma}{\Delta}{\copowerterm{t}{u}}{\Ccopower{\VA}{\CB}}
\endprooftree
\qquad
\prooftree
\Cj{\Gamma}{\Delta}{s}{\Ccopower{\VA}{\CB}}
\quad
\Cj{\Gamma, \, \In{x}{\VA}}{\In{y}{\CB}}{t}{\CC}
\justifies
\Cj{\Gamma}{\Delta}{\copowerlet{x}{y}{s}{t}}{\CC}
\endprooftree
\\[\gspace]
\prooftree
\Cj{\Gamma}{\Delta}{t}{\Czero}
\justifies
\Cj{\Gamma}{\Delta}{\Cimage{t}}{\CA}
\endprooftree
\qquad
\prooftree
\Cj{\Gamma}{\Delta}{t}{\CA}
\justifies
\Cj{\Gamma}{\Delta}{\Cinl{t}}{\CA \Cplus \CB}
\endprooftree
\qquad
\prooftree
\Cj{\Gamma}{\Delta}{t}{\CB}
\justifies
\Cj{\Gamma}{\Delta}{\Cinr{t}}{\CA \Cplus \CB}
\endprooftree
\\[\gspace]
\prooftree
\Cj{\Gamma}{\Delta}{s}{\CA \Cplus \CB}
\quad
\Cj{\Gamma}{\In{x}{\CA}}{t}{\CC}
\quad
\Cj{\Gamma}{\In{y}{\CB}}{u}{\CC}
\justifies
\Cj{\Gamma}{\Delta}{\Ccase{s}{x}{t}{y}{u}}{\CC}
\endprooftree
\\[\gspace]
\prooftree
\Cj{\Gamma}{\In{z}{\CA}}{t}{\CB}
\justifies
\Vj{\Gamma}{\llam{z}{\CA}{t}}{\CA \lpop \CB}
\endprooftree
\qquad
\prooftree
\Vj{\Gamma}{s}{\CA \lpop \CB} 
 \quad
\Cj{\Gamma}{\Delta}{t}{\CA} 
\justifies
\Cj{\Gamma}{\Delta}{\lappl{s}{t}}{\CB}
\endprooftree
\end{gather*}
\caption{Typing rules for the enriched effect calculus}
\label{figure:effects:typing}
\vspace*{30pt}
\end{figure}

\begin{figure}  % [t]
\vspace*{30pt}
\begin{align*}
& \Veq{\Gamma}{t}{\Vstar}{\Vone} & & \text{if $\Vj{\Gamma}{t}{\Vone}$} 
\\
& \Veq{\Gamma}{\Vfst{\Vpair{t}{u}}}{t}{\VA} 
  && \text{if $\Vj{\Gamma}{t}{\VA}$ and $\Vj{\Gamma}{u}{\VB}$} 
\\
& \Veq{\Gamma}{\Vsnd{\Vpair{t}{u}}}{u}{\VB} 
  && \text{if $\Vj{\Gamma}{t}{\VA}$ and $\Vj{\Gamma}{u}{\VB}$} 
\\
& \Veq{\Gamma}{\Vpair{\Vfst{t}}{\Vsnd{t}}}{t}{\VA \Vprod \VB} 
  && \text{if $\Vj{\Gamma}{t}{\VA \Vprod \VB}$}
\\
& \Veq{\Gamma}{\Vappl{(\Vlam{x}{\VA}{t})}{u}}{t[u/x]}{\VB}
  && \text{if $\Vj{\Gamma,\In{x}{\VA}}{t}{\VB}$ and $\Vj{\Gamma}{u}{\VA}$}
\\
& \Veq{\Gamma}{\Vlam{x}{\VA}{(\Vappl{t}{x})}}{t}{\VA \Vfun \VB} 
  && \text{if $\Vj{\Gamma}{t}{\VA \Vfun \VB}$ and $x \not\in \Gamma$}
\\
& \Ceq{\Gamma}{\Delta}{t}{\Cstar}{\Cone} & & \text{if $\Cj{\Gamma}{\Delta}{t}{\Cone}$} 
\\
& \Ceq{\Gamma}{\Delta}{\Cfst{\Cpair{t}{u}}}{t}{\CA} 
  && \text{if $\Cj{\Gamma}{\Delta}{t}{\CA}$ and $\Cj{\Gamma}{\Delta}{u}{\CB}$} 
\\
& \Ceq{\Gamma}{\Delta}{\Csnd{\Cpair{t}{u}}}{u}{\CB} 
  && \text{if $\Cj{\Gamma}{\Delta}{t}{\CA}$ and $\Cj{\Gamma}{\Delta}{u}{\CB}$} 
\\
& \Ceq{\Gamma}{\Delta}{\Cpair{\Cfst{t}}{\Csnd{t}}}{t}{\CA \Cprod \CB} 
  && \text{if $\Cj{\Gamma}{\Delta}{t}{\CA \Cprod \CB}$}
\\
& \Ceq{\Gamma}{\Delta}{\Cappl{(\Clam{x}{\VA}{t})}{u}}{t[u/x]}{\CB}
  && \text{if $\Cj{\Gamma,\In{x}{\VA}}{\Delta}{t}{\CB}$ and $\Vj{\Gamma}{u}{\VA}$}
\\
& \Ceq{\Gamma}{\Delta}{\Clam{x}{\VA}{(\Cappl{t}{x})}}{t}{\VA \Cfun \CB} 
  && \text{if $\Cj{\Gamma}{\Delta}{t}{\VA \Cfun \CB}$ and $x \not\in \Gamma, \Delta$}
\\
& \Veq{\Gamma}{\Ilet{\Itop}{t}}{t}{\CA}
  && \text{if $\Vj{\Gamma}{t}{\CA}$}
\\
& \Ceq{\Gamma}{\Delta}{\Ilet{t}{u[\Itop/x]}}{u[t / x]}{\CA}
  && \text{if $\Cj{\Gamma}{\Delta}{t}{\CI}$ and $\Cj{\Gamma}{\In{x}{\CI}}{u}{\CA}$}
\\
& \Veq{\Gamma}{\banglet{x}{\bang{t}}{u}}{u[t/x]}{\CB}
  && \text{if $\Vj{\Gamma}{t}{\VA}$ and $\Vj{\Gamma,\, \In{x}{\VA}}{u}{\CB}$}
\\
& \Ceq{\Gamma}{\Delta}{\banglet{x}{t}{u[\bang{x}/y]}}{u[t / y]}{\CB}
  && \text{if $\Cj{\Gamma}{\Delta}{t}{\Cbang{\VA}}$ and $\Cj{\Gamma}{\In{y}{\Cbang{\VA}}}{u}{\CB}$}
\\
& \Ceq{\Gamma}{\Delta}{\copowerlet{x}{y}{\copowerterm{t}{s}}{u}}{u[t,\! s / x, \! y]}{\CC} 
 && \text{if $\Vj{\Gamma}{t}{\VA}$, $\; \Cj{\Gamma}{\Delta}{s}{\CB}$, and}
\\
&
 && \quad\text{$\Cj{\Gamma, \In{x}{\VA}}{\In{y}{\CB}}{u}{\CC}$}
\\
& \Ceq{\Gamma}{\Delta}{\copowerlet{x}{y}{t}{u[\copowerterm{x}{y} / z]}}{u[t / z]}{\CC} 
&& \text{if $\Cj{\Gamma}{\Delta}{t}{\Ccopower{\VA}{\CB}}$ and $\Cj{\Gamma}{\In{z}{\Ccopower{\VA}{\CB}}}{u}{\CC}$}
\\
& \Ceq{\Gamma}{\Delta}{\Cimage{t}}{u[t/x]}{\CA}
 && \text{if $\Cj{\Gamma}{\Delta}{t}{\Czero}$ and $\Cj{\Gamma}{\In{x}{\Czero}}{u}{\CA}$}
\\
& {\Gamma} \! \mid \! {\Delta} \vdash {\Ccase{\Cinl{t}}{x}{u}{y}{u'}}
 && \text{if $\Cj{\Gamma}{\In{x}{\CA}}{u}{\CC}$ and $\Cj{\Gamma}{\In{y}{\CB}}{u'}{\CC}$}
\\
& \phantom{\Gamma\mid \Delta \vdash \quad} =  {u[t/x]} \, \colon \, {\CC}
 && \quad\text{and $\Cj{\Gamma}{\Delta}{t}{\CA}$}
\\
& {\Gamma} \! \mid \! {\Delta}  \vdash {\Ccase{\Cinr{t}}{x}{u}{y}{u'}} 
 && \text{if $\Cj{\Gamma}{\In{x}{\CA}}{u}{\CC}$ and $\Cj{\Gamma}{\In{y}{\CB}}{u'}{\CC}$}
\\
& \phantom{\Gamma\mid \Delta \vdash \quad} = {u'[t/y]} \, \colon \, {\CC}
&& \quad \text{and $\Cj{\Gamma}{\Delta}{t}{\CB}$}
\\
& \Gamma  \! \mid  \! \Delta \! \vdash \! \Ccase{t}{x}{u[\Cinl{x} / z]}{y}{ u[\Cinr{y} / z]} \hspace*{-50pt}\\
& \phantom{\Gamma\mid \Delta \vdash \quad} = u[t / z] \, \colon \, \CC
 && \text{if $\Cj{\Gamma}{\Delta}{t}{\CA \Cplus \CB}$ and $\Cj{\Gamma}{\In{z}{\CA \Cplus \CB}}{u}{\CC}$}
\\
& \Ceq{\Gamma}{\Delta}{\lappl{(\llam{x}{\CA}{t})}{u}}{t[u/x]}{\CB} 
  && \text{if $\Cj{\Gamma}{\In{x}{\CA}}{t}{\CB}$ and $\Cj{\Gamma}{\Delta}{u}{\CA}$}
\\
& \Veq{\Gamma}{\llam{x}{\CA}{(\lappl{t}{x})}}{t}{\CA \lpop \CB}
 && \text{if $\Vj{\Gamma}{t}{\CA \lpop \CB}$ and $x\notin \Gamma$}
\end{align*}
\caption{Equality rules for the enriched effect calculus}
\label{figure:effects:equalities}
\vspace*{30pt}
\end{figure}

The enriched effect calculus has two typing judgements:
\begin{align*}
\text{(i)} \; \; & \Vj{\Gamma}{t}{\VB} 
& 
\text{(ii)} \; \; & \Cj{\Gamma}{\In{z}{\CA}}{t}{\CB} \enspace ,
\end{align*}
where $\Gamma$ is a context of value-type assignments to variables. On the right of $\Gamma$ is
a \emph{stoup}, which may either be empty, as in the case of judgement (i), or may
consist of a unique type assignment $\In{z}{\CA}$, in which case the type on the right of the
turnstyle is also required to be a computation type, as in (ii).
The typing rules are given in Figure~\ref{figure:effects:typing}. In them, $\Delta$ ranges over an
arbitrary (possibly empty) stoup, and the rules are only applicable in the case
of typing judgements that conform to (i) or (ii) above. 

\begin{prop}[Weakening]
\leavevmode
\label{prop:weak}
If $\Cj{\Gamma}{\Delta}{t}{\CA}$  and variable $x$ is not contained in $\Gamma,\Delta$
then $\Cj{\Gamma,\In{x}{\CB}}{\Delta}{t}{\CA}$.
\end{prop}
\begin{prop}[Substitution]
\label{prop:subst}
\leavevmode
\begin{enumerate}[\em(1)]
\item \label{subst:i}
If $\Cj{\Gamma, \In{x}{\VA}}{\Delta}{t}{\VB}$ and  $\Vj{\Gamma}{u}{\VA}$ and 
then $\Cj{\Gamma}{\Delta}{t[u/x]}{\VB}$.
\item \label{subst:ii}
If $\Cj{\Gamma}{\In{x}{\CA}}{t}{\CB}$ and $\Cj{\Gamma}{\Delta}{u}{\CA}$ then
then $\Cj{\Gamma}{\Delta}{t[u/x]}{\CB}$.
\end{enumerate}
\end{prop}

\noindent
A simple consequence of the propositions above is that EEC satisfies the
``shift'' property: if $\Cj{\Gamma}{\In{x}{\CA}}{t}{\CB}$ then 
$\Vj{\Gamma, \In{x}{\CA}}{t}{\CB}$. 
See~\cite{EMSb} for further discussion of syntactic properties of
EEC.
%(If $\Cj{\Gamma}{\In{x}{\CA}}{t}{\CB}$ then
%$\Cj{\Gamma, \In{y}{\CA}}{\In{x}{\CA}}{t}{\CB}$ by Proposition~\ref{prop:weak}.
%Since $\Vj{\Gamma, \In{x}{\CA}}{y}{\CA}$, we then have
%$\Vj{\Gamma, \In{y}{\CA}}{t[y/x]}{\CB}$   by Proposition~\ref{prop:subst}.\label{subst:ii}.
%Whence $\Vj{\Gamma, \In{x}{\CA}}{t}{\CB}$, by a similar argument
%using  Proposition~\ref{prop:weak} and 
%Proposition~\ref{prop:subst}.\label{subst:i}.)

Rules for equalities between typed terms are presented in
Figure~\ref{figure:effects:equalities}. They are to be considered in
addition to the expected (typed) congruence and $\alpha$-equivalence rules.
The equations of Figure~\ref{figure:effects:equalities} have been formulated in
such a way that the smallest $\alpha$-equivalence-respecting congruence 
containing these equalities is automatically closed under 
the substitution operations of Proposition~\ref{prop:subst}.

% Note that the enriched effect calculus can be seen as a fragment of Barber and Plotkin's Dual 
% Intuitionistic/Linear Lambda calculus (DILL)~\cite{Barber:97}.

The relationship between the enriched effect calculus and other calculi is discussed in
detail in~\cite{EMSb}. We summarise the main points relevant to the present paper.

The fragment of EEC obtained by removing the type constructors $\VA \Vfun \VB$,
$\,\CA \lpop \CB$, 
$\, \Ccopower{\VA}{\CB}$, $\, \Czero$ and $\CA \Cplus \CB$ is called the
\emph{effect calculus} (EC) in~\cite{EMSb}.\footnote{This differs mildly from the ``effect calculus''
of~\cite{EMSb} through not having value-type function spaces.} The effect calculus is
equivalent to Levy's CBPV (with complex stacks, finitary syntax version) modulo
the difference that CBPV has one further type constructor:
value-type sums. Since, on the one hand, value-type sums can be easily added to the 
effect calculus~\cite{EMS}, and,
on the other, just as easily removed from CBPV, we consider this difference as minor.
Thus it seems fair to view the effect calculus (where value-type sums can be included
if desired) as, essentially,  a reformulation of CBPV using a syntax and
presentation influenced 
by linear logic. 
In particular,
the style of typing rule we have given owes a debt to 
Barber and Plotkin's Dual Intuitionistic Linear Logic~\cite{Barber:97}.
The influence of linear logic is, of course, even  more apparent
in the case of the enriched effect calculus.
In~\cite{EMS,EMSc}, it is shown that EEC is a conservative extension
of EC, thus the presence of the additional linear primitives does not alter the 
properties of the core type constructors from EC.

It is also natural to compare EEC with ILL. In the present paper, we do this
informally and crudely.\footnote{A less crude comparison retains the distinction between
computation and value type, and compares with Benton's mixed linear/non-linear
logic~\cite{Benton:95}, in which a similar distinction is maintained.
Such a comparison produces identical results: the translation is sound, but
neither complete nor full.}
We include EEC in ILL by ignoring the distinction between value and computation
types, and mapping all type constructors to their evident (mainly synonymous)
linear counterparts. For example, both $\Vfun$ and $\Cfun$ get mapped to
the intuitionistic function space of ILL; both $\Vprod$ and $\Cprod$ get mapped to
the linear ``with'' $\&$; both $\Vone$ and $\Cone$ get mapped to the unit of
the intuitionistic ``with'', which is usually denoted $\top$; and $\CI$ gets mapped
to the unit of the linear tensor, which is usually denoted $1$. This translation
from EEC to ILL is ``sound'' in the sense that terms that are equal in EEC get mapped to
equal terms in ILL. (This is a consequence of the simple observation that the typing rules
and equations of EEC are all have direct counterparts in 
the presentation of ILL of~\cite{Barber:97}.) However, the translation
is not ``complete'': 
terms of the same type whose translations are equal in ILL need not be equal
in EEC. It is also not ``full'', there exist terms in ILL whose type lies in the
EEC fragment of ILL, but which are not equal to the translation of any EEC term.

\section{Call-by-value and call-by-name translations into EEC}
\label{section:cbv:cbn}

There is a standard call-by-value translation
of typed $\lambda$-calculus into Moggi's
computational metalanguage~\cite{Moggi:91}, 
Filinski's effect PCF~\cite{Filinski:phd}, and Levy's
CBPV~\cite{Levy:book}. Similarly, there is a 
standard call-by-name translation into the latter two, which
exploits the existence of computation types.\footnote{Moggi~\cite{Moggi:91} 
and Benton and Wadler~\cite{BW:96} refer to a different
``lazy''  translation as call-by-name.}
We recall these translations
using the syntax of the enriched effect calculus. 
% As in CBPV we can define a call-by-value and a call-by-name translation of simply typed lambda calculus into $\ECE$. 

\begin{figure}  % [t]
\begin{gather*}
\prooftree
\justifies
\Lj{\Theta,\, \In{x}{\sigma}}{x}{\sigma}
\endprooftree
\qquad
\prooftree
\justifies 
\Vj{\Theta}{\Lstar}{\Lone}
\endprooftree
\\[\gspace]
\prooftree
\Lj{\Theta}{M}{\sigma}
  \quad
\Lj{\Theta}{N}{\tau}
\justifies 
\Lj{\Theta}{\Lpair{M}{N}}{\sigma \Lprod \tau}
\endprooftree
\qquad
\prooftree
\Lj{\Theta}{M}{\sigma \Vprod \tau}
\justifies 
\Lj{\Theta}{\Lfst{M}}{\sigma}
\endprooftree
\qquad
\prooftree
\Lj{\Theta}{M}{\sigma \Lprod \tau}
\justifies 
\Lj{\Theta}{\Lsnd{M}}{\tau}
\endprooftree
\\[\gspace]
\prooftree
\Lj{\Theta,\,\In{x}{\sigma}}{M}{\tau}
\justifies
\Lj{\Theta}{\Llam{x}{\sigma}{M}}{\sigma \Lfun \tau}
\endprooftree
\qquad
\prooftree
\Lj{\Theta}{M}{\sigma \Lfun \tau} 
  \quad
\Lj{\Theta}{N}{\sigma} 
\justifies
\Lj{\Theta}{\Lappl{M}{N}}{\tau}
\endprooftree
\end{gather*}
\caption{Typing rules for simply-typed $\lambda$-calculus}
\label{figure:lambda:typing}
\end{figure}

As a source calculus, we use the simply-typed $\lambda$-calculus with types $\sigma, \tau, \dots$
given by:
\[
\sigma \: ::= \: \alpha \, \mid \, \Lone \, \mid \, \sigma \Lprod \tau \, \mid \, \sigma \Lfun \tau \enspace ,
\]
where $\alpha$ ranges over a collection of type constants.
We use $\Theta$ to range over finite contexts 
$\In{x_1}{\sigma_1}, \ldots, \In{x_n}{\sigma_n}$, and $M,N$ to range over terms
of the simply-typed $\lambda$-calculus, using the 
syntax given by the typing rules in Figure~\ref{figure:lambda:typing}.

\begin{figure}[t]
\begin{align*}
\cbv{\alpha} & = \alpha & \cbn{\alpha} & = \comptype{\alpha} \\
\displaybreak[0]
\cbv{\Lone} & = \Vone & \cbn{\Lone} & = \Cone \\
\displaybreak[0]
\cbv{(\sigma \Lprod \tau)} & = \cbv{\sigma} \Vprod \cbv{\tau}
 &  \cbn{(\sigma \Lprod \tau)} & = \cbn{\sigma} \Cprod \cbn{\tau} \\
\displaybreak[0]
\cbv{(\sigma \Lfun \tau)} & = \cbv{\sigma} \Vfun {\Cbang{(\cbv{\tau})}}
 &  \cbn{(\sigma \Lfun \tau)} & = \cbn{\sigma} \Cfun \cbn{\tau} \enspace .
\end{align*}
\caption{Cbv and cbn translations of simply-typed $\lambda$-calculus}
\label{figure:cbv:cbn}
\end{figure}

The call-by-value interpretation translates a type $\sigma$ to
a value type $\cbv{\sigma}$. The 
call-by-name interpretation translates it to
a computation type $\cbn{\sigma}$. Both translations
are defined in Figure~\ref{figure:cbv:cbn}.
For the translations of type constants,
we assume that each type constant $\alpha$ of the typed $\lambda$-calculus,
is included as a value-type constant in EEC, and has an associated 
computation-type constant $\comptype{\alpha}$. 
Note that the definition of $\cbv{(\sigma \Lfun \tau)}$ could equally well have been
given as $\cbv{\sigma} \Cfun {\Cbang{\cbv{\tau}}}$, which, considered as
a value type, is isomorphic to the gven translation. Our reason for instead choosing 
$\cbv{\sigma} \Vfun {\Cbang{(\cbv{\tau})}}$ is that this simplifies 
the statement of Theorem~\ref{theorem:recover:cbv} below.

On terms, the cbv translation maps a judgement 
$\Lj{\In{x_1}{\sigma_1}, \ldots, \In{x_n}{\sigma_n}}{M}{\tau}$ to
\[\Vj{\In{x_1}{\cbv{\sigma_1}}, \ldots, \In{x_n}{\cbv{\sigma_n}}}{\cbv{M}}{\Cbang{\cbv{\tau}}} \enspace .\]
It is inductively defined by:
\begin{align*}
\cbv{x} \: & = \: \bang{x} \\
\cbv{\Lstar} \: & = \: \bang{\Vstar} \\
\cbv{\Lpair{M}{N}} \: & = \: \banglet{x}{\cbv{M}}{\banglet{y}{\cbv{N}}{\bang{\Vpair{x}{y}}}} \\
\cbv{(\Lfst{M})} \: & = \: \banglet{z}{\cbv{M}}{\bang{\Vfst{z}}} \\
\cbv{(\Lsnd{M})} \: & = \: \banglet{z}{\cbv{M}}{\bang{\Vsnd{z}}} \\
\cbv{(\Llam{x}{\sigma}{M})} \: & = \: \bang{(\Vlam{x}{\cbv{\sigma}}{\cbv{M}})} \\
\cbv{(\Lappl{M}{N})} \: & = \: \banglet{f}{\cbv{M}}{\banglet{x}{\cbv{N}}{\Vappl{f}{x}}} \enspace .
\end{align*}
The cbn translation
maps a judgement 
$\Lj{\In{x_1}{\sigma_1}, \ldots, \In{x_n}{\sigma_n}}{M}{\tau}$ to
\[\Vj{\In{x_1}{\cbn{\sigma_1}}, \ldots, \In{x_n}{\cbn{\sigma_n}}}{\cbn{M}}{\cbn{\tau}} \enspace ,\]
and simply uses the constructs associated with the computation-type constructors
$\Cone$, $\Cprod$ and $\Cfun$ to mimic the corresponding constructs for $\Lone$, $\Lprod$ and $\Lfun$
in the simply-typed $\lambda$-calculus. Since this is essentially trivial, we omit the details.

The call-by-value and call-by-name translations into EEC induce equational theories
on simply-typed $\lambda$-terms. In the case of call-by-value, the resulting equational
theory is that of Moggi's \emph{computational $\lambda$-calculus}, $\lambda_c$,~\cite{Moggi:89}.
In the case of call-by-name, it is the usual $\beta\eta$-equality theory.
The propositions below state this formally, and also assert that the translations
into EEC are \emph{full} in the sense that every EEC term of translated type
is equal to the translation of a simply-typed term. In the statements, and henceforth,
we  write $\LCeq{\Theta}{M}{N}{\tau}$ for equality in Moggi's $\lambda_c$, and
$\BEeq{\Theta}{M}{N}{\tau}$ for $\beta\eta$-equality.
\begin{prop}[Soundness and full  completeness of $\cbv{(\cdot)}$]
\leavevmode
\label{prop:cbv} % {prop:cbpv}
\begin{enumerate}[\em(1)]
\item \label{cbv:i} % {cbpv:i}
If $\LCeq{\Theta}{M}{N}{\tau}$ then 
$\Veq{\cbv{\Theta}}{\cbv{M}}{\cbv{N}}{\Cbang{\,\cbv{\tau}}}$.

\item \label{cbv:ii} % {cbpv:ii}
If $\Lj{\Theta}{M,N}{\tau}$ and
$\Veq{\cbv{\Theta}}{\cbv{M}\!}{\cbv{N}\!}{\Cbang{\cbv{\tau}}}$ then
$\LCeq{\Theta}{M\!}{N\!}{\tau}$.

\item \label{cbv:iii}
If $\Vj{\cbv{\Theta}\!}{t}{\Cbang{\cbv{\tau}}}$ then there
exists a term $\Lj{\Theta\!}{M}{\tau}$ such that 
$\Veq{\cbv{\Theta}\!}{\cbv{M}}{t}{\Cbang{\,\cbv{\tau}}}$.
\end{enumerate}
\end{prop}
\noindent
Here, statement~\ref{cbv:i} asserts soundness, statement~\ref{cbv:ii} completeness,
and statement~\ref{cbv:iii} fullness.

\begin{prop}[Soundness and full  completeness of $\cbn{(\cdot)}$]
\leavevmode
\label{prop:cbn}
\begin{enumerate}[\em(1)]

\item \label{cbn:i}
If 
$\BEeq{\Theta}{M}{N}{\tau}$ then
$\Veq{\cbn{\Theta}}{\cbn{M}}{\cbn{N}}{\cbn{\tau}}$.

\item \label{cbn:ii}
If $\Lj{\Theta}{M,N}{\tau}$ and  $\Veq{\cbn{\Theta}}{\cbn{M}}{\cbn{N}}{\cbn{\tau}}$
then $\BEeq{\Theta}{M}{N}{\tau}$.

\item \label{cbn:iii}
If $\Vj{\cbn{\Theta}\!}{t}{\cbn{\tau}}$ then there
exists a term $\Lj{\Theta\!}{M}{\tau}$ such that 
$\Veq{\cbn{\Theta}\!}{\cbn{M}}{t}{\cbn{\tau}}$.
\end{enumerate}
\end{prop}

\proof[Outline proof of Propositions~\ref{prop:cbv} and~\ref{prop:cbn}]
The call-by-value and call-by-name translations 
of typed $\lambda$-calculus into Levy's CBPV are
known to be fully complete~\cite[Appendix A]{Levy:book}. These translations 
thus transfer to the \emph{effect calculus} (EC) of~\cite{EMSb}, which is essentially
equivalent to CBPV. The resulting translations into EC are essentially
identical to those given above, modulo the inclusion of EC in the
enriched effect calculus. This inclusion is shown to be fully
complete in~\cite{EMS,EMSc}.
\qed

\noindent
The repeat appearance of the word ``essentially'' in the outline proof above calls for
clarification. As already discussed in Section~\ref{section:calculus},
the equivalence between CBPV and the effect calculus requires choosing
the  correct version of CBPV (with complex stacks and finitary syntax), and
ignoring the fact that CBPV has value-type sums but  EC does not. Anyway, such
issues are a distraction here, since the translations do not involve sum types,
and Levy's proofs of full completeness transfer directly to EC. Second,
the call-by-value translation we have given into EEC
is not literally identical to the translation into EC. The difference is that,
in the case of EC (as defined in~\cite{EMSb}, see Section~\ref{section:calculus}),
one has to define 
$\cbv{(\sigma \! \Lfun \!\tau)} = \cbv{\sigma} \! \Cfun \! {\Cbang{\cbv{\tau}}}$, because
value-type function space is not available. This difference is, however, trivial
since the two function spaces are isomorphic as value types.

Via the inclusion of EEC as a fragment of ILL, the translations defined above can also be viewed as
translations into ILL. In the case of call-by-value, the resulting translation into ILL is 
exactly Benton and Wadler's call-by-value translation from~\cite{BW:96}. As emphasised in
\emph{op.\ cit.}, this translation is not complete relative to $\LCequals$ because it enforces the \emph{commutativity}
of effects. For example, the two terms below,
\begin{align}
\label{Mone}
\Lj{\In{f}{\Lone \Lfun \Lone },\, \In{g}{\Lone \Lfun \Lone} & }{\Lappl{\Lappl{(\Llam{x}{\Lone}{\Llam{y}{\Lone}{\Lstar}})}{(\Lappl{f}{\Lstar})}}{(\Lappl{g}{\Lstar})}}{\Lone}
\\
\label{Mtwo}
\Lj{\In{f}{\Lone \Lfun \Lone },\, \In{g}{\Lone \Lfun \Lone} & }{\Lappl{\Lappl{(\Llam{x}{\Lone}{\Llam{y}{\Lone}{\Lstar}})}{(\Lappl{g}{\Lstar})}}{(\Lappl{f}{\Lstar})}}{\Lone} \enspace ,
\end{align}
which are not equated by $\LCequals$, are equated by the translation.
It is also known that the call-by-value translation into ILL is not full~\cite{Hasegawa:Flops:02}.

\section{Linearly-used continuations in EEC}
\label{sec:lin:cps}

In~\cite{Plotkin}, Plotkin gave continuation passing style (CPS) translations 
of call-by-value and call-by-name $\lambda$-calculi into the 
$\lambda$-calculus. As emerged from the work of Moggi~\cite{Moggi:89,Moggi:91},
the typed version of Plotkin's call-by-value translation is
sound relative to the equational theory, $\LCequals$, of the computational $\lambda$-calculus.
Although Plotkin's original
call-by-name translation validates only the $\beta$-law, a variation
due to Reus and Streicher~\cite{RS} is sound for $\BEequals$.
A feature shared by all these translations is that the usage of  continuations
within them is linear. This aspect has been formalized by Hasegawa.
In~\cite{Hasegawa:Flops:02}, he studies a 
call-by-value translation from typed $\lambda$-calculus
into intuitionistic linear type theory (ILL) in which the types 
of the translation enforce the linear usage of continuations. In essence,
this translation is Plotkin's original call-by-value translation, but 
carried out within a  linear typing discipline.
In~\cite{Hasegawa:Flops:04}, Hasegawa gives a  corresponding linear version
of the (Reus-Streicher) call-by-name CPS translation.
Although Hasegawa's translations
are into ILL, one sees straightforwardly that they land
inside the EEC fragment of ILL.\footnote{Actually, in~\cite{Hasegawa:Flops:04}, Hasegawa 
gives a call-by-name translation for a variant of Parigot's $\lambda\mu$-calculus~\cite{Parigot:92}
extending typed $\lambda$-calculus. The full translation goes outside of EEC. Here, we consider
just the translation restricted to typed $\lambda$-calculus, which does land in EEC.}
We now recall these translations, defining them directly  as translations into EEC.

The call-by-value interpretation translates a type $\sigma$
to a value type $\cbvLincps{\sigma}$, and the call-by-name interpretation
translates $\sigma$ to a computation type $\cbnLincps{\sigma}$, as defined in
Figure~\ref{figure:lincps}.
As is standard for CPS translations, 
they are defined relative to the choice of a 
 ``result'' type, $\CR$.
Using EEC as the target language, 
it is essential that $\CR$ be a computation type, otherwise
the translations would not produce legal types.
Unless specified otherwise, we let $\CR$ be an arbitrary but fixed computation type. 
However, we shall often need to specify otherwise. As will be seen,
many results below will work in two special cases only:
when $\CR$ is either a computation-type constant or the type $\CI$.

We remark that the combination of function-space constructs that appears in the call-by-value translation
of $\sigma \Lfun \tau$, in Figure~\ref{figure:lincps},
is forced by the desire to ensure that continuations are
linearly used. The linear usage itself is implemented by selecting
$\lpop$ for the right-hand arrow. This, in turn, requires the 
computation-type arrow $\Cfun$ to be used in the 
type $\cbvLincps{\tau} \Cfun \CR$, which types continations.
The left-hand arrow is then forced to be $\Vfun$ since its
codomain $(\cbvLincps{\tau} \Cfun \CR) \lpop  \CR$ is a value type.
(It is possible to reduce the number of different
function-space constructs that appear in the definition of $\cbvLincps{(\sigma \Lfun \tau)}$
to two. For example, one could define $\cbvLincps{(\sigma \Lfun \tau)}$
to be $(\cbvLincps{\tau} \Cfun \CR) \lpop  (\cbvLincps{\sigma} \Cfun  \CR)$, which is
isomorphic to the definition of Figure~\ref{figure:lincps}. Another possibility is to reformulate
EEC using a single type constructor to implement both value-type and computation-type
function spaces, as in the conference version of this 
paper~\cite{EMS:fossacs}. However,
both these alternatives have the disadvantage, compared with the route we have taken,
of complicating the results of
Sections~\ref{section:canonical} and~\ref{section:recovering}.

\begin{figure*}[t] % [htbp]
\begin{align*}
\cbvLincps{\alpha} \: & = \: \alpha & 
\cbnLincps{\alpha} \: & = \: \comptype{\alpha} \\
\cbvLincps{\Lone} \: & = \: \Vone & 
\cbnLincps{\Lone} \: & = \: \Czero \\
\cbvLincps{(\sigma \Lprod \tau)} \: & = \: \cbvLincps{\sigma} \Vprod \cbvLincps{\tau} &  
\cbnLincps{(\sigma \Lprod \tau)} \: & = \: \cbnLincps{\sigma} \Cplus \cbnLincps{\tau} \\
\cbvLincps{(\sigma \Lfun \tau)} \: & = \: \cbvLincps{\sigma} \Vfun ((\cbvLincps{\tau} \Cfun \CR) \lpop  \CR)
 &  \cbnLincps{(\sigma \Lfun \tau)} \: & = \: \Ccopower{(\cbnLincps{\sigma}  \lpop \CR)}{\cbnLincps{\tau}} \enspace .
\end{align*}
\caption{Cbv and cbn linear-use CPS translations of typed $\lambda$-calculus.}
\label{figure:lincps}
\end{figure*}

For a typing context $\Theta = \In{x_1}{\sigma_1}, \ldots, \In{x_n}{\sigma_n}$, define
\[
\cbvLincps{\Theta} \: = \: \In{x_1}{\cbvLincps{\sigma_1}}, \ldots, \In{x_n}{\cbvLincps{\sigma_n}} \enspace .
\]
Then the cbv translation on terms~\cite{Hasegawa:Flops:02} maps a judgement 
$\Lj{\Theta}{M}{\tau}$ to
\[\Vj{\cbvLincps{\Theta}}{\cbvLincps{M}}{(\cbvLincps{\tau} \Cfun \CR) \lpop \CR} \enspace ,\]
It is defined inductively by (using the typings of Figure~\ref{figure:lambda:typing}):
\begin{align*}
\cbvLincps{x} \: & = \: \llam{k}{\cbvLincps{\sigma} \Cfun \CR}{\,\Cappl{k}{x}} \\
\cbvLincps{\Lstar} \: & = \: \llam{k}{\Vone \Cfun \CR}{\,\Cappl{k}{\Vstar}} \\
\cbvLincps{\Lpair{M}{N}} \: & = \: \llam{k}{(\cbvLincps{\sigma} \Lprod \cbvLincps{\tau}) \Cfun \CR}
     {\,\lappl{\cbvLincps{M}}{\Clam{x}{\cbvLincps{\sigma}}
         {\,\lappl{\cbvLincps{N}}{\Clam{y}{\cbvLincps{\tau}}{\,\Cappl{k}{\Vpair{x}{y}}}}}}} \\
\cbvLincps{(\Lfst{M})} \: & = \: \llam{k}{\cbvLincps{\sigma} \Cfun \CR}
     {\,\lappl{\cbvLincps{M}}{\Clam{z}{\cbvLincps{\sigma} \times \cbvLincps{\tau}}{\, \Cappl{k}{\Vfst{z}}}}} \\                  
\cbvLincps{(\Lsnd{M})} \: & = \:  \llam{k}{\cbvLincps{\tau} \Cfun \CR}
     {\,\lappl{\cbvLincps{M}}{\Clam{z}{\cbvLincps{\sigma} \times \cbvLincps{\tau}}{\, \Cappl{k}{\Vsnd{z}}}}} \\ 
\cbvLincps{(\Llam{x}{\sigma}{M})} \: & = \: \llam{k}{(\cbvLincps{\sigma} \Vfun (\cbvLincps{\tau} \Cfun \CR) \lpop \CR) \Cfun \CR}
     {\, \Cappl{k}{\Vlam{x}{\cbvLincps{\sigma}{\,\cbvLincps{M}}}}} \\
\cbvLincps{(\Lappl{M}{N})} \: & = \: \llam{k}{\cbvLincps{\tau} \Cfun \CR}
     {\,\lappl{\cbvLincps{M}}{\Clam{f}{\cbvLincps{\sigma} \Vfun (\cbvLincps{\tau} \Cfun \CR) \lpop \CR}
         {\,\lappl{\cbvLincps{N}}{\Clam{x}{\cbvLincps{\sigma}}{\,\lappl{\Vappl{f}{x}}{k}}}}}} 
\end{align*}

Similarly, define
\[
\cbnLincps{\Theta} \lpop \CR \: = \: \In{x_1}{\cbnLincps{\sigma_1} \lpop \CR}, \ldots, \In{x_n}{\cbnLincps{\sigma_n} \lpop \CR} \enspace .
\]
The cbn translation~\cite{Hasegawa:Flops:04}
maps a typing judgement $\Lj{\Theta}{M}{\tau}$, as above, to
\[\Vj{\cbnLincps{\Theta} \lpop \CR}{\cbnLincps{M}}{\cbnLincps{\tau} \lpop \CR} \enspace .
\]
Its inductive definition is given by:
\begin{align*}
\cbnLincps{x} \: & = \: x \\
\cbnLincps{\Lstar} \: & = \: \llam{k}{\Czero}{\Cimage{k}} \\
\cbnLincps{\Lpair{M}{N}} \: & = \: \llam{k}{\cbvLincps{\sigma} \Cplus \cbvLincps{\tau}}
     {\,\Ccase{k}{x}{\,\lappl{\cbnLincps{M}\,}{x}}{\,y}{\,\lappl{\cbnLincps{N}\,}{y}}} \\
\cbnLincps{(\Lfst{M})} \: & = \: \llam{k}{\cbnLincps{\sigma}}
     {\,\lappl{\cbnLincps{M}\,}{\Cinl{k}}} \\
\cbnLincps{(\Lsnd{M})} \: & = \: \llam{k}{\cbnLincps{\tau}}
     {\,\lappl{\cbnLincps{M}\,}{\Cinr{k}}} \\
\cbnLincps{(\Llam{x}{\sigma}{M})} \: & = \: \llam{k}{\Ccopower{(\cbnLincps{\sigma} \lpop \CR)}{\cbnLincps{\tau}}}
     {\, \copowerlet{x}{h}{k}{\, \lappl{\cbnLincps{M}\,}{h}}} \\
\cbnLincps{(\Lappl{M}{N})} \: & = \: \llam{k}{\cbnLincps{\tau}}
     {\,\lappl{\cbnLincps{M}\,}{\copowerterm{(\cbnLincps{N})\, }{\, k}}}
\end{align*}

The results below list the properties we shall establish 
of the two translations. Proofs will be given in Section~\ref{section:recovering}.

\begin{prop}[Soundness of $\cbvLincps{(\cdot)}$]
\label{prop:cbvLincps:sound}
If $\LCeq{\Theta}{M}{N}{\tau}$ then
$\Veq{\cbvLincps{\Theta}}{\cbvLincps{M}}{\cbvLincps{N}}{(\cbvLincps{\tau} \Cfun \CR) \lpop \CR}$.
\end{prop}

\begin{prop}[Soundness of $\cbnLincps{(\cdot)}$]
\label{prop:cbnLincps:sound}
If $\BEeq{\Theta}{M}{N}{\tau}$ then
$\Veq{\cbnLincps{\Theta}\lpop \CR}{\cbnLincps{M}}{\cbnLincps{N}}{\cbnLincps{\tau} \lpop \CR}$.
\end{prop}

\begin{thm}[Full completeness of $\cbvLincps{(\cdot)}$]
\label{thm:cbvLincps:complete}
Suppose $\CR$ is either: (i) a computation-type constant, or (ii) the type $\CI$. Then:
\begin{enumerate}[\em(1)]
\item \label{fcl:i}
If $\Lj{\Theta}{M,N}{\tau}$ and 
$\Veq{\cbvLincps{\Theta}}{\cbvLincps{M}\!}{\!\cbvLincps{N}}{(\cbvLincps{\tau} \!\Cfun \!\CR) \lpop \CR}$ then
$\LCeq{\Theta}{M\!}{\!N}{\tau}$.

\item \label{fcl:ii}
If $\Vj{\cbvLincps{\Theta}}{t}{(\cbvLincps{\tau} \Cfun \CR) \lpop \CR}$  then there
exists a term $\Lj{\Theta}{M}{\tau}$ such that 
$\Veq{\cbvLincps{\Theta}}{\cbvLincps{M}}{t}{(\cbvLincps{\tau} \Cfun \CR) \lpop \CR}$.
\end{enumerate}
\end{thm}

\begin{thm}[Full completeness of $\cbnLincps{(\cdot)}$]
\label{thm:cbnLincps:complete}
Suppose $\CR$ is either: (i) a computation-type constant different from $\comptype{\alpha}$, for every
simply-typed $\lambda$-calculus type constant $\alpha$; or (ii) the type $\CI$. Then:
\begin{enumerate}[\em(1)]
\item If $\Lj{\Theta}{M,N}{\tau}$ and
$\Veq{\cbnLincps{\Theta}\lpop \CR}{\cbnLincps{M}}{\cbnLincps{N}}{\cbnLincps{\tau} \lpop \CR}$
then $\BEeq{\Theta}{t}{u}{\tau}$.

\item If $\Vj{\cbnLincps{\Theta} \lpop \CR}{t}{\cbnLincps{\tau}\lpop \CR}$ then there
exists a term $\Lj{\Theta}{M}{\tau}$ such that 
$\Veq{\cbnLincps{\Theta} \lpop \CR}{\cbnLincps{M}}{t}{\cbnLincps{\tau}\lpop \CR}$.
\end{enumerate}
\end{thm}

Theorems~\ref{thm:cbvLincps:complete} and \ref{thm:cbnLincps:complete} are analogous to
full completeness results obtained by  Hasegawa for the linear-use CPS translations into
ILL. In~\cite{Hasegawa:Flops:02} he proves 
full completeness for the call-by-value linear-use CPS translation of Moggi's
computational $\lambda$-calculus~\cite{Moggi:89} into ILL.
A similar result holds for the call-by-name translation
of~\cite{Hasegawa:Flops:04} restricted to the simply-typed $\lambda$-calculus
(private communication). In both cases, Hasegawa considers translations in
which $\CR$ is taken to be a computation-type constant. 

We remark that, in the case that $\CR$ is a computation-type constant,
Theorems~\ref{thm:cbvLincps:complete} and \ref{thm:cbnLincps:complete} follow as a consequence of 
Hasegawa's full completeness results for the translations into ILL.
This is because, even though the inclusion of EEC in ILL is neither complete (faithful) nor full,
it is sound (preserves equalities)~\cite{EMSb}. Hence, for any fully complete
translation into ILL that  factors through this inclusion, such as
the linear-use CPS translations, the 
factoring translation into EEC is also fully complete. A little thought shows 
that a
similar style of argument cannot be used to derive Hasegawa's results as a consequence of 
Theorems~\ref{thm:cbvLincps:complete} and \ref{thm:cbnLincps:complete}.
Thus, full completeness with respect to ILL seems a strictly stronger property
than full completeness with respect to EEC. 
Nevertheless, even though Theorems~\ref{thm:cbvLincps:complete} and \ref{thm:cbnLincps:complete},
in the case that $\CR$ is a computation-type constant, follow from Hasegawa's results
(and not vice-versa), our method of proof is different, and of interest in its own right --- see below.

Furthermore, Theorems~\ref{thm:cbvLincps:complete} and \ref{thm:cbnLincps:complete} extend Hasegawa's
result in a different direction. They apply also when the type $\CI$ is used for $\CR$. 
In the case of the call-by-value translation, this property distinguishes
between the translations into EEC and ILL. Indeed, the call-by-value linear-use CPS
translation into ILL is not complete if $\CI$ is used for $\CR$. A simple counterexample
is given by the two $\lambda$-calculus terms (\ref{Mone}) and (\ref{Mtwo}),
which translate to terms:
\begin{align*}
\Lj{\In{f}{\cbvLincps{(\Lone \Lfun \Lone)}},\, \In{g}{\cbvLincps{(\Lone \Lfun \Lone)}} & }
{\cbvLincps{(\Lappl{\Lappl{(\Llam{x}{\Lone}{\Llam{y}{\Lone}{\Lstar}})}{(\Lappl{f}{\Lstar})}}{(\Lappl{g}{\Lstar})})}}
{\Cbang{(\cbvLincps{\Lone})}}
\\
\Lj{\In{f}{\cbvLincps{(\Lone \Lfun \Lone)}},\, \In{g}{\cbvLincps{(\Lone \Lfun \Lone)}} & }
{\cbvLincps{(\Lappl{\Lappl{(\Llam{x}{\Lone}{\Llam{y}{\Lone}{\Lstar}})}{(\Lappl{g}{\Lstar})}}{(\Lappl{f}{\Lstar})})}}
{\Cbang{(\cbvLincps{\Lone})}} \enspace .
\end{align*}
Noting that $\cbvLincps{(\Lone \Lfun \Lone)} = \Vone \Vfun ((\Vone \Cfun \CI) \lpop \CI)$,
which is isomorphic,
in EEC and (hence) in ILL, to $\CI$;
and $\Cbang{(\cbvLincps{\Lone})} = \Cbang{\Vone}$, which is also isomorphic to $\CI$, on can calculuate
that the two translated terms are transported along these isomorphisms to:
\begin{align*}
\Vj{\In{f}{\CI},\, \In{g}{\CI} & }{\Ilet{f}{\Ilet{g}{\Itop}}}{\CI}
\\
\Vj{\In{f}{\CI},\, \In{g}{\CI} & }{\Ilet{g}{\Ilet{f}{\Itop}}}{\CI} \enspace .
\end{align*}
These terms are equal in ILL but not in EEC. (This is reminiscent of the fact that the
cbv translation $\cbv{(\cdot)}$  of Section~\ref{section:cbv:cbn}, when taken into ILL,
enforces the commutativity of effects~\cite{BW:96}; but not identical, because 
$\cbvLincps{(\cdot)}$ is not, in general, isomorphic to $\cbv{(\cdot)}$.)

%It is worth observing that the call-by-name translation is, in a certain sense,
%almost as trivial as the call-by-name translation of Section~\ref{section:cbv:cbn}. [[BLAH]]

Our proof of Theorems~\ref{thm:cbvLincps:complete} and \ref{thm:cbnLincps:complete}
goes via factoring the $\cbvLincps{(\cdot)}$ and $\cbnLincps{(\cdot)}$ through a 
single generic linear-use CPS translation of the entire
enriched effect calculus into itself. This translation, which is the main contribution 
of the paper,  is presented in the next section.

\section{Generic linear-use CPS self-translation of EEC}
\label{section:canonical}

\begin{figure}[h!]
\vspace*{10pt}
\begin{align*}
\CpsVT{\VconstA} & \: = \: \VconstA  &
\CpsCT{\CconstA} & \: = \: \begin{cases} \CconstA & \text{if $\CconstA \neq \CR$} \\
                                          \CI & \text{if $\CconstA = \CR$} \end{cases}
\\
\CpsVT{\Vone} & \: = \: \Vone &
\CpsCT{\Cone} & \: = \: \Czero \\
\CpsVT{(\VA \Vprod \VB)} & \: = \: \CpsVT{\VA} \Vprod \CpsVT{\VB} &
\CpsCT{(\CA \Cprod \CB)} & \: = \:  \CpsCT{\CA} \Cplus \CpsCT{\CB} \\
\CpsVT{(\VA \Vfun \VB)} & \: = \: \CpsVT{\VA} \Vfun \CpsVT{\VB} &
\CpsCT{(\VA \Cfun \CB)} & \: = \: {\Ccopower{(\CpsVT{\VA})}{\CpsCT{\CB}}} \\
\CpsVT{\CA} & \: = \: \CpsCT{\CA} \lpop \CR &
\CpsCT{\CI} & \: = \: {\CR} \\
\CpsVT{(\CA \lpop \CB)} & \: = \:  \CpsCT{\CB} \lpop \CpsCT{\CA}  &
\CpsCT{(\Cbang{\VA})} & \: =\:  \CpsVT{\VA} \Cfun  \CR \\
& &
\CpsCT{(\Ccopower{\VA}{\CB})} & \: = \: \CpsVT{\VA} \Cfun \CpsCT{\CB} \\
& &
\CpsCT{\Czero} & \: = \: \Cone \\
& &
\CpsCT{(\CA \Cplus \CB)} & \: = \: \CpsCT{\CA} \Cprod  \CpsCT{\CB} 
%  \\ && \CpsCT{\CR} & \: = \: {\CI}
\end{align*}
\caption{Linear-use CPS translation of EEC types.}
\label{fig:cps:type}
\vspace*{10pt}
%\end{figure}
%\begin{figure}[htbp]
\begin{align*}
\CpsCT{z}  & \: = \:  k_z
\\
\CpsCT{\Cstar}  & \: = \:  \Cimagetype{\CD}{k_z}
\\
\CpsCT{\Cpair{t}{u}}  & \: = \:  
   {\Ccase{k_z}{k_x}{\CpsCT{t}\, [k_x/k_z]}{k_y}{\CpsCT{u}\,[k_y/k_z]}} 
\\
\CpsCT{\Cfst{t}}  & \: = \:  
   \CpsCT{t}\, [\Cinl{k_z}/k_z]
\\
\CpsCT{\Csnd{t}}  & \: = \:  
   \CpsCT{t}\, [\Cinr{k_z}/k_z]
\\
\CpsCT{(\Clam{x}{\VA}{t})}  & \: = \:  
   \copowerlet{x}{h}{k_z}{\,\CpsCT{t}\,[h/k_z]} 
\\
\CpsCT{(\Cappl{s}{t})}  & \: = \:  
   \CpsCT{s}\,[\copowerterm{(\CpsVT{t})}{k_z} \, / \, k_z]
\\
\CpsCT{(\Ilet{t}{u})}  & \: = \:  
   \CpsCT{t}\,[\lappl{\CpsVT{u}\,}{k_z} \, / \, k_z]
\\
\CpsCT{(\banglet{x}{t}{u})}  & \: = \:  
   \CpsCT{t}\,[(\Clam{x}{\CpsVT{\VA}}{\,\lappl{\CpsVT{u}\,}{k_z}})\, / \, k_z]
\\
\CpsCT{(\copowerterm{t}{u})}  & \: = \:  
   \CpsCT{u}\,[\Cappl{k_z}{\CpsVT{t}} \, / \, k_z]
\\
\CpsCT{(\copowerlet{x}{y}{s}{t})}  & \: = \:  
   \CpsCT{s}\,[(\Clam{x}{\CpsVT{\VA}}{\,\CpsCT{t}\,[k_z/k_y]})\, / \, k_z] 
\\
\CpsCT{(\Cimage{t})}  & \: = \:  \CpsCT{t}\,[\Cstar \, / \, k_z]
\\
\CpsCT{(\Cinl{t})}  & \: = \:  \CpsCT{t}\,[\Cfst{k_z} \, / \, k_z]
\\
\CpsCT{(\Cinr{t})}  & \: = \:  \CpsCT{t}\,[\Csnd{k_z} \, / \, k_z]
\\ 
\CpsCT{(\Ccase{s}{x}{t}{y}{u})}  & \: = \:  
   \CpsCT{s}\,[\Cpair{\CpsCT{t}\,[k_z/k_x]}{\,\CpsCT{u}\,[k_z/k_y]} \, / \, k_z] 
\\
\CpsCT{(\lappl{s}{t})}  & \: = \:  
   \CpsCT{t}\,[\lappl{\CpsVT{s}\,}{k_z} \, / \, k_z]
\end{align*}
\caption{Linear-use CPS translation of computation terms.}
\label{fig:cps:compterms}
\vspace*{10pt}
\end{figure}
\begin{figure}[t]
\begin{align*}
\CpsVT{x}  & \: = \:  x 
\\
\CpsVT{\Vstar}  & \: = \:  \Vstar 
\\
\CpsVT{\Vpair{t}{u}}  & \: = \:  \Vpair{\CpsVT{t}}{\CpsVT{t}}
\\
\CpsVT{(\Vfst{t})}  & \: = \:  \Vfst{\CpsVT{t}}
\\
\CpsVT{(\Vsnd{t})}  & \: = \:  \Vsnd{\CpsVT{t}}
\\
\CpsVT{(\Vlam{x}{\VA}{t})} & \: = \:  \Vlam{x}{\VA}{\,\CpsVT{t}}
\\
\CpsVT{(\Vappl{t}{u})}  & \: = \:  \Vappl{\CpsVT{t}\,}{\CpsVT{u}}
\\
\CpsVT{\Cstar}  & \: = \:  \llam{k}{\Czero}{\,\Cimagetype{\CR}{k}}
\\
\CpsVT{\Cpair{t}{u}}  & \: = \:  \llam{k}{\CpsCT{\CA} \Cplus \CpsCT{\CB}}
   {\,\Ccase{k}{k_x}{\lappl{\CpsVT{t}\,}{k_x}}{k_y}{\lappl{\CpsVT{u}\,}{k_y}}} 
\\
\CpsVT{\Cfst{t}}  & \: = \:  \llam{k}{\CpsCT{\CA}}
   {\,\lappl{\CpsVT{t}\,}{\Cinl{k}}}
\\
\CpsVT{\Csnd{t}}  & \: = \:  \llam{k}{\CpsCT{\CB}}
   {\,\lappl{\CpsVT{t}\,}{\Cinr{k}}}
\\
\CpsVT{(\Clam{x}{\VA}{t})}  & \: = \:  \llam{k}{\Ccopower{\CpsVT{\VA}}{\CpsCT{\CB}}}
   {\,\copowerlet{x}{h}{k}{\,\lappl{\CpsVT{t}\,}{h}}}
\\
\CpsVT{(\Cappl{s}{t})}  & \: = \:  \llam{k}{\CpsCT{\CB}}
   {\,\lappl{\CpsVT{s}\,}{\copowerterm{(\CpsVT{t})}{k}}}
\\
\CpsVT{\Itop}  & \: = \:  \llam{k}{\CR}{\, k}
\\
\CpsVT{(\Ilet{t}{u})}  & \: = \:  \llam{k}{\CpsCT{\CA}}{\, 
   \lappl{\CpsVT{t}\,}{\lappl{\CpsVT{u}\,}{k}}}
\\
\CpsVT{(\bang{t})}  & \: = \:  \llam{k}{\CpsVT{\VA} \Cfun \CR}
   {\,\Cappl{k\,}{\CpsVT{t}}}
\\
\CpsVT{(\banglet{x}{t}{u})}  & \: = \:  \llam{k}{\CpsCT{\CB}}
   {\,\lappl{\CpsVT{t}\,}{\Clam{x}{\CpsVT{\VA}}{\,\lappl{\CpsVT{u}\,}{k}}}}
\\
\CpsVT{(\copowerterm{t}{u})}  & \: = \:   \llam{k}{\CpsVT{\VA} \Cfun \CpsCT{\CB}}
   {\,\lappl{\CpsVT{u}\,}{\Cappl{k\,}{\CpsVT{t}}}} 
\\
\CpsVT{(\copowerlet{x}{y}{s}{t})}  & \: = \:  \llam{k}{\CpsCT{\CC}}
   {\,\lappl{\CpsVT{s}\,}{\Clam{x}{\CpsVT{\VA}}{\,\CpsCT{t}\,[k/k_y]}}} 
\\
\CpsVT{(\Cimage{t})}  & \: = \:  \llam{k}{\CpsCT{\CA}}
   {\,\lappl{\CpsVT{t}\,}{\Cstar}}
\\ 
\CpsVT{(\Cinl{t})}  & \: = \:  \llam{k}{\CpsCT{\CA} \Cprod \CpsCT{\CB}}
   {\,\lappl{\CpsVT{t}\,}{\Cfst{k}}}
\\
\CpsVT{(\Cinr{t})}  & \: = \:  \llam{k}{\CpsCT{\CA} \Cprod \CpsCT{\CB}}
   {\,\lappl{\CpsVT{t}\,}{\Csnd{k}}}
\\ 
\CpsVT{(\Ccase{s}{x}{t}{y}{u})}  & \: = \:   \llam{k}{\CpsCT{\CC}}
   {\,\lappl{\CpsVT{s}\,}{\Cpair{\CpsCT{t}\,[k/k_x]}{\,\CpsCT{u}\,[k/k_y]}}} 
\\
\CpsVT{(\llam{z}{\CA}{t})}  & \: = \:  \llam{k}{\CpsCT{\CB}}{\,\CpsCT{t}\,[k/k_z]}
\\ 
\CpsVT{(\lappl{s}{t})}  & \: = \:  \llam{k}{\CpsCT{\CB}}
   {\,\lappl{\CpsVT{t}\,}{\lappl{\CpsVT{s}\,}{k}}}
\end{align*}
\caption{Linear-use CPS translation of value terms.}
\label{fig:cps:valterms}
\end{figure}

The generic linear-use  CPS translation, from  EEC 
to itself,  maps a
value type $\VA$ to a value type $\CpsVT{\VA}$ and a computation
type $\CA$ to a computation type $\CpsCT{\CA}$, as defined in
Figure~\ref{fig:cps:type}. Note that the translation of a computation type $\CA$ 
as a computation type, $\CpsCT{\CA}$, 
is defined prior to its translation as a value type, $\CpsVT{\CA}$.
Note also that, in the case that the result type $\CR$ is
a computation-type constant, it is given special treatment. Otherwise 
it is translated in the same way as any other type. This means that, when
$\CR$ is either a computation-type constant or $\CI$, we obtain the
complementary equations $\CpsCT{\CR} = \CI$ and $\CpsCT{\CI} = \CR$,
exhibiting the computation types $\CR$ and $\CI$ as a \emph{dual pair}. 
Other examples of dual pairs are:
$\Cone$ and $\Czero$; $\, \CA \Cprod \CB$ and $\CA \Cplus \CB$;
$\,\VA \Cfun \CB$ and $\Ccopower{\VA}{\CB}$; and $\CconstA$ (for 
$\CconstA \neq \CR$) with itself. Thus the
only computation types without a  dual (in this simple sense) 
are those of the form $\Cbang{\CA}$. The reason that
such dual pairs arise in the translation is that the  translation acts
\emph{contravariantly} on computation types, in a sense which will
be made clear below, but which is already implicit in the
identity $\CpsVT{(\CA \lpop \CB)} = \CpsCT{\CB} \lpop \CpsCT{\CA}$. For this reason,
each computation type is translated to a computation type that possesses the 
dual universal property to its own. The contravariance
of the computation-type translation also underlies
the identity $\CpsVT{\CA} = \CpsCT{\CA} \lpop \CR$, which ``negates'' the
computation-type translation of a computation type in order to bring it into
the \emph{covariant} world of value-type translations. 
We remark that in the conference version of this paper~\cite{EMS:fossacs}, this identity 
held only up to isomorphism, leading to syntactic
complications. The implementation of the equality as a syntactic identity, in
Figure~\ref{fig:cps:type}, is possible in the present paper, 
because we distinguish between  value-type and computation-type products and
between value- and computation-type function spaces.

%The computation types
%include three dual pairs: (i)  $\algone$ and $\algzero$, (ii)
%$\CA \prodtype \CB$ and  $\CA \algplus \CB$, and (iii)
%$\VA \to \CB$ and $\ltensortype{\VA}{\CB}$, each of which
%has its members swapped by application of the
%translation $\CpsCT{(\cdot)}$ on computation types. 

To define the translation of  terms, we translate a typing judgement 
$\Vj{\Gamma}{t}{\VA}$ as:
\[
\Vj{\CpsVT{\Gamma}}{\CpsVT{t}}{\CpsVT{\VA}} \enspace ,
\]
where $\CpsVT{\Gamma}$ is the context obtained by applying $\CpsVT{(-)}$ to every type in $\Gamma$. 
A typing judgement $\Cj{\Gamma}{\In{z}{\CA}}{t}{\CB}$ is translated to:
\[
\Cj{\CpsVT{\Gamma}}{\In{k_z}{\CpsCT{\CB}}}{\CpsCT{t}}{\CpsCT{\CA}} \enspace .
\]
The change of direction here is the contravariance we referred to above.
The translations are given in Figures~\ref{fig:cps:valterms} 
and~\ref{fig:cps:compterms} respectively.
In these figures, 
each line corresponds to one of the typing rules
in Figure~\ref{figure:effects:typing}, and the type 
and term names are taken from these rules. 
Observe that each typing rule that mentions $\Delta$ 
has two cases:  one, in Figure~\ref{fig:cps:valterms},
for  empty stoup in Figure~\ref{fig:cps:compterms}, and one 
for non-empty stoup.
Also note that, in Figure~\ref{fig:cps:compterms},
we always use $\In{z}{\CD}$ for the content of a non-empty stoup called $\Delta$ in 
Figure~\ref{figure:effects:typing}. We remark that,
because we have the identity $\CpsVT{(\CA \lpop \CB)} = \CpsCT{\CB} \lpop \CpsCT{\CA}$,
the translations are simpler than those
given in the conference version of the paper~\cite{EMS:fossacs}, which involved
specified isomorphisms in lieu of the identity.

The remainder of the section is devoted to establishing properties of the self-translation.
As a first observation, we note that if $\Vj{\Gamma}{t}{\VA}$, where $x$ is not
contained in $\Gamma$, then the terms, appearing in each of the translated judgements
(cf.\ Proposition~\ref{prop:weak})
\[
\Vj{\CpsVT{\Gamma}}{\CpsVT{t}}{\CpsVT{\VA}}
\qquad
\Vj{\CpsVT{\Gamma}, \In{x}{\CpsVT{\VB}}}{\CpsVT{t}}{\CpsVT{\VA}} \enspace ,
\]
are identical (as the notation suggests). Similarly, if
$\Cj{\Gamma}{\In{z}{\CA}}{t}{\CB}$, where again $x$ is not in $\Gamma$,
then the two terms
\[
\Cj{\CpsVT{\Gamma}}{\In{k_z}{\CpsCT{\CB}}}{\CpsCT{t}}{\CpsCT{\CA}} 
\qquad
\Cj{\CpsVT{\Gamma}, \In{x}{\CpsVT{\VC}}}{\In{k_z}{\CpsCT{\CB}}}{\CpsCT{t}}{\CpsCT{\CA}} 
\]
are identical. These observations are easily seen to hold by 
a straightforward induction on the structure of $t$.

The interaction between the self-translation and substitution is more subtle.
Each of the two cases of Proposition~\ref{prop:subst} splits into
two subcases, one for empty $\Delta$, and one for non-empty $\Delta$,
resulting in the four cases considered in the proposition below. 
\begin{prop}[Substitution]
\leavevmode
\label{prop:trans:subs}
\begin{enumerate}[\em(1)]
\item \label{subs:i}
If $\Vj{\Gamma, \In{x}{\VA}}{t}{\VB}$ and $\Vj{\Gamma}{u}{\VA}$ then
$\Veq{\CpsVT{\Gamma}}{\CpsVT{(t[u/x])}}{\CpsVT{t}\, [\CpsVT{u}\, / \, x]}{\CpsVT{\VB}}$.
\item \label{subs:ii}
If $\Cj{\Gamma, \In{x}{\VA}}{\In{z}{\CD}}{t}{\CB}$ and $\Vj{\Gamma}{u}{\VA}$ then
\[\Ceq{\CpsVT{\Gamma}}{\In{k_z}{\CpsCT{\CB}}}
   {\CpsCT{(t[u/x])}}{\CpsCT{t}\, [\CpsVT{u}\, / \, x]}{\CpsCT{\CD}} \enspace .\]
\item \label{subs:iii}
If $\Cj{\Gamma}{\In{x}{\CA}}{t}{\CB}$ and $\Vj{\Gamma}{u}{\CA}$
then \[\Veq{\CpsVT{\Gamma}}{\CpsVT{(t[u/x])}}
    {\llam{k}{\CpsCT{\CB}}{\,\lappl{\CpsVT{u}\,}{\CpsCT{t}\,[k\,/\,k_x]}}}{\CpsCT{\CB} \lpop \CR} \enspace .\]
\item \label{subs:iv}
If $\Cj{\Gamma}{\In{x}{\CA}}{t}{\CB}$ and $\Cj{\Gamma}{\In{z}{\CD}}{u}{\CA}$ 
then \[\Ceq{\CpsVT{\Gamma}}{\In{k_z}{\CpsCT{\CB}}}{\CpsCT{(t[u/x])}}
    {\CpsCT{u}\, [(\CpsCT{t}\,[k_z\, / k_x])\, / \, k_z]}{\CpsCT{\CD}} \enspace .\]
\end{enumerate}
\end{prop}
\proof
By induction on $t$.

Statements \ref{subs:i} and \ref{subs:ii} are proved simultaneously. For example,
if $t$ is $\llam{z}{\CB_1}{t'}$, where $\CB$ is $\CB_1 \lpop \CB_2$, then
statement \ref{subs:i} applies, and we must show that 
$\CpsVT{((\llam{z}{\CB_1}{t'})[u/x])} = \CpsVT{(\llam{z}{\CB_1}{t'})}\, [\CpsVT{u}\, / \, x]$.
The induction hypothesis, given by statement \ref{subs:ii}, is
$\CpsCT{(t'[u/x])} = \CpsCT{(t')}\, [\CpsVT{u}\, / \, x]$. And indeed:
\begin{align*}
\CpsVT{((\llam{z}{\CB_1}{t'})[u/x])} \: 
&  = \: \CpsVT{(\llam{z}{\CB_1}{\, t'[u/x]})}
\\
& = \: \llam{\,k_z}{\CpsCT{\CB_2}}{\, \CpsCT{(t'[u/x])}}
\\
& = \: \llam{\,k_z}{\CpsCT{\CB_2}}{\, \CpsCT{(t')}\, [\CpsVT{u}\, / \, x]}
  && \text{by induction hypothesis}
\\
& = \: \CpsVT{(\llam{z}{\CB_1}{t'})}\, [\CpsVT{u}\, / \, x] \enspace .
\end{align*}

We illustrate the proof of statement \ref{subs:iii} in the case that $t$ is
$\Ccase{t'}{y}{t_1}{z}{t_2}$, where
$\Cj{\Gamma}{\In{x}{\CA}}{t'}{\CC_1 \oplus \CC_2}$, and
$\Cj{\Gamma}{\In{y}{\CC_1}}{t_1}{\CB}$, and
$\Cj{\Gamma}{\In{z}{\CC_2}}{t_2}{\CB}$. Then:
\begin{align*}
\CpsVT{(t[u/x])} \: 
& = \: \CpsVT{(\Ccase{t'[u/x]}{y}{t_1}{z}{t_2})}
\\
& = \:   \llam{k}{\CB}{\, 
   \lappl{\CpsVT{(t'[u/x])}\,}{\Cpair{\CpsCT{t_1}\, [k/k_y]}{\CpsCT{t_2}\, [k/k_z]}}}
\\
& = \:  \llam{k}{\CB}{\, 
      \lappl{\CpsVT{u}\,}{\CpsCT{(t')}\,[\Cpair{\CpsCT{t_1}\, [k/k_y]}{\CpsCT{t_2}\, [k/k_z]}\,/\,k_x]}}
  && \text{by induction hypothesis}
\\
& = \: \llam{k}{\CB}{\, 
          \lappl{\CpsVT{u}\,}{\CpsCT{(\Ccase{t'}{y}{t_1}{z}{t_2})}\, [k/k_x]}}
\\
& =  \: \llam{k}{\CpsCT{\CB}}{\,\lappl{\CpsVT{u}\,}{\CpsCT{t}\,[k/k_x]}}
\enspace .
\end{align*}

We omit the proof of statement \ref{subs:iv}, which is straightforward.
\qed

We now have the machinery necessary to establish the first of the main properties
of the self-translation, its equational soundness.

\begin{thm}[Soundness]
\label{thm:cps:soundness}
\leavevmode
\begin{enumerate}[\em(1)]
\item If $\Veq{\Gamma}{t}{u}{\VA}$ then 
$\Veq{\CpsVT{\Gamma}}{\CpsVT{t}}{\CpsVT{u}}{\CpsVT{\VA}}$.

\item If $\Ceq{\Gamma}{\In{z}{\CA}}{t}{u}{\CB}$
then $\Ceq{\CpsVT{\Gamma}}{\In{k_z}{\CpsCT{\CB}}}{\CpsCT{t}}{\CpsCT{u}}{\CpsCT{\CA}}$.
\end{enumerate}
\end{thm}

\proof
Define $\Vsim{\Gamma}{t}{u}{\VA}$  to hold if 
$\Veq{\CpsVT{\Gamma}}{\CpsVT{t}}{\CpsVT{u}}{\CpsVT{\VA}}$,
and similarly $\Csim{\Gamma}{\In{z}{\CA}}{t}{u}{\CB}$ to hold if 
$\Ceq{\CpsVT{\Gamma}}{\In{k_z}{\CpsCT{\CB}}}{\CpsCT{t}}{\CpsCT{u}}{\CpsCT{\CA}}$.
Trivially, $\sim$ is a type-respecting equivalence relation.
By the compositional definition of $\CpsVT{(\cdot)}$ and $\CpsCT{(\cdot)}$
it is an $\alpha$-equivalence respecting congruence. It remains to
verify that $\sim$ satisfies the equalities of 
Figure~\ref{figure:effects:equalities}. Once again, every equality in
which $\Delta$ appears, splits into two cases, one for empty $\Delta$, and one for non-empty
$\Delta$. This means that  the 24 equalities of Figure~\ref{figure:effects:equalities},
give rise to 39 equalities that need verifying. We consider two cases, by way of illustration.

For the first case, suppose $\Vj{\Gamma}{t}{\VA}$ and $\Vj{\Gamma,\, \In{x}{\VA}}{u}{\CB}$.
We show that
\[
\Veq{\CpsVT{\Gamma}}{\CpsVT{(\banglet{x}{\bang{t}}{u})}}{\CpsVT{(u[t/x])}}{\CpsCT{\CB} \lpop \CR} \enspace .
\]
For this,
\begin{align*}
\CpsVT{(\banglet{x}{\bang{t}}{u})} \:
&  = \: \llam{k}{\CpsCT{\CB}}
   {\,\lappl{\CpsVT{(\bang{t})}\,}{\Clam{x}{\CpsVT{\VA}}{\,\lappl{\CpsVT{u}\,}{k}}}}
\\
& = \:
\llam{k}{\CpsCT{\CB}}
   {\,\Cappl{(\Clam{x}{\CpsVT{\VA}}{\,\lappl{\CpsVT{u}\,}{k}})\,}{\CpsVT{t}}}
 && \text{def.~of $\CpsVT{(\bang{t})}$}
\\
& = \:\llam{k}{\CpsCT{\CB}}
   {\,\lappl{\CpsVT{u}\,[\CpsVT{t}\,/\, x]\,}{k}}
 && \text{$\beta$ equality}
\\
& = \: {\CpsVT{u}\,[\CpsVT{t}\,/\, x]\,}
 && \text{$\eta$ equality}
\\
& = \: \CpsVT{(u[t/x])}
 && \text{Prop.~\ref{prop:trans:subs}.\ref{subs:i}} \enspace .
\end{align*}

For the second case, 
suppose $\Cj{\Gamma}{\In{z}{\CD}}{t}{\Cbang{\VA}}$ and $\Cj{\Gamma}{\In{y}{\Cbang{\VA}}}{u}{\CB}$.
We show that 
\[\Ceq{\Gamma}{\In{k_z}{\CpsCT{\CB}}}{\CpsCT{(\banglet{x}{t}{u[\bang{x}/y]})}}{\CpsCT{(u[t / y])}}{\CpsCT{\CD}}
\enspace .\]
For this,
\begin{align*}
\CpsCT{(\banglet{x}{t}{u[\bang{x}/y]})} \:
& = \: \CpsCT{t}\,[(\Clam{x}{\CpsVT{\VA}}{\,\lappl{\CpsVT{(u[\bang{x}/y])}\,}{k_z}})\, / \, k_z]
\\
& = \: \CpsCT{t}\,[(\Clam{x}{\CpsVT{\VA}}{\,\lappl{\CpsVT{(\bang{x})}\,}{\CpsCT{u} \,[k_z/k_y]}})\, / \, k_z]
 && \text{Prop.~\ref{prop:trans:subs}.\ref{subs:iii}}
\\
& = \: \CpsCT{t}\,[(\Clam{x}{\CpsVT{\VA}}{\, \Cappl{(\CpsCT{u} \,[k_z/k_y])}{\CpsVT{x}}})
\, / \, k_z]
 && \text{def.~of $\CpsVT{(\bang{x})}$}
\\
& = \: \CpsCT{t}\,[(\Clam{x}{\CpsVT{\VA}}{\, \Cappl{(\CpsCT{u} \,[k_z/k_y])}{x}})
\, / \, k_z]
 && \text{def.~of $\CpsVT{x}$}
\\
& = \: \CpsCT{t}\,[(\CpsCT{u} \,[k_z/k_y]) \, / \, k_z]
 && \text{$\eta$ equality}
\\
& = \: \CpsCT{(u[t / y])}
 && \text{Prop.~\ref{prop:trans:subs}.\ref{subs:iv}} \enspace .
\end{align*}
We comment that the second step above, employs the equality
\[\CpsVT{(u[\bang{x}/y])} \: = \:
\llam{k}{\CpsCT{\CB}}{\,\lappl{\CpsVT{(\bang{x})}\,}{\CpsCT{u}\,[k\,/\,k_y]}} \enspace ,
\]
whose strict derivation from Proposition~\ref{prop:trans:subs}.\ref{subs:iii} 
invokes the coincidence of the two terms:
\[\Cj{\CpsVT{\Gamma}}{\In{k_y}{\CB}}{u}{\CpsVT{\VA} \Cfun \CR} 
\qquad
\Cj{\CpsVT{\Gamma}, \In{x}{\CpsVT{\VA}}}{\In{k_y}{\CB}}{u}{\CpsVT{\VA} \Cfun \CR} 
\enspace .\]
Having made this point once, we shall not comment further on such small issues arising from weakening.
\qed

We now come to the central result of the paper:
if $\CR$ is either a computation-type constant or $\CI$ then 
the self-translation is involutive up to isomorphism (Theorem~\ref{theorem:involution}).
That is, the translation of the translation of a term is equal,
modulo type isomorphism, to the original term.
To state the involution property, we first define the required isomorphisms.
For each value type $\VA$, we define a closed EEC term, $\Viso{\VA} \colon \CpsVVT{\VA} \Vfun \VA$,
for which there exists a corresponding closed
term 
$\Viso{\VA}^{-1} \colon \VA \Vfun \CpsVVT{\VA}$ 
such that the equations
$\Vlam{x}{\VA}{\,\Vappl{\Viso{\VA}}{\Vappl{\Viso{\VA}^{-1}}{x}}} = \Vlam{x}{\VA}{x}$ and
$\Vlam{x}{\CpsVVT{\VA}}{\,\Vappl{\Viso{{\VA}^{-1}}}{\Vappl{\Viso{\VA}}{x}}} = \Vlam{x}{\CpsVVT{\VA}}{x}$
hold in the EEC equational theory.
Similarly, for each computation type $\CA$, we define a closed EEC term 
$\Ciso{\CA} \colon \CpsCCT{\CA} \lpop \CA$ that is a linear isomorphism. That is,
the inverse is given by a closed term $\Ciso{\CA}^{-1} \colon \CA \lpop \CpsCCT{\CA}$
such that the equations asserting the mutual inverse properties again hold.
The families of terms $\Viso{\VA}$ and $\Ciso{\CA}$ are mutually defined by induction on 
their types in Figure~\ref{fig:tiip}. Note that, for a computation type $\CA$, the
linear isomorphism $\Ciso{\CA}$ is defined first, and the definition of $\Viso{\CA}$ 
depends on it.
Note also that the clauses for function types require the inverses
of previously defined terms, which, since they are inverses, are 
uniquely determined up to provable equality. Their existence 
is assured by
the lemma below, which therefore establishes
that Figure~\ref{fig:tiip} is a good definition.

\begin{figure}
\begin{align*}
% \CpsVVT{\alpha}  \: & = \: \alpha  & 
\Viso{\alpha} \: & = \: \Vlam{x}{\alpha}{x} 
\\
% \CpsVVT{\Vone}  \: & = \: \Vone & 
\Viso{\Vone} \: & =  \: \Vlam{x}{\Vone}{\Vstar} 
\\
% \CpsVVT{(\VA \Vprod \VB)}  \: & = \: \CpsVVT{\VA} \Vprod \CpsVVT{\VB} & 
\Viso{\VA \Vprod \VB} \: & = \: \Vlam{z}{\CpsVVT{\VA} \Vprod \CpsVVT{\VB}}
      {\, \Vpair{\Vappl{\Viso{\VA}}{\Vfst{z}}}{\Vappl{\Viso{\VB}}{\Vsnd{z}}}}
\\
\Viso{\VA \Vfun \VB} \: & = \: \Vlam{f}{\CpsVVT{\VA} \Vfun \CpsVVT{\VB}}
      {\, \Vlam{x}{\VA}{\, \Vappl{\Viso{\VB}}{\Vappl{f}{\Vappl{\Viso{\VA}^{-1}}{x}}}}}
\\
\Viso{\CA} \: & = \: \Vlam{h}{\CI \lpop \CpsCCT{\CA}}
      {\, \lappl{\Ciso{\CA}\,}{\lappl{h\,}{\Itop}}}
\\
\Viso{\CA \lpop \CB} \: & = \: \Vlam{h}{\CpsCCT{\CA} \lpop \CpsCCT{\CB}}
      {\, \llam{x}{\CA}{\,\lappl{\Ciso{\CB}\,}{\lappl{h\,}{\lappl{\Ciso{\CA}^{-1}\,}{x}}}}}
\\[5pt]
\Ciso{\CconstA} \: & = \: \llam{z}{\CconstA}{z} 
\\
\Ciso{\Cone} \: & = \: \llam{z}{\Cone}{\Cstar} 
\\
\Ciso{\CA \Cprod \CB} \: & = \: \llam{z}{\CpsCCT{\CA} \Cprod \CpsCCT{\CB}}
      {\, \Cpair{\lappl{\,\Ciso{\CA}\,}{\Cfst{z}}}{\lappl{\,\Ciso{\CB}\,}{\Csnd{z}}\,}}
\\
\Ciso{\VA \Cfun \CB} \: & = \: \llam{f}{\CpsVVT{\VA} \Cfun \CpsCCT{\CB}}
      {\, \Clam{x}{\VA}{\, \lappl{\Ciso{\CB}\,}{\Cappl{f}{\Vappl{\Viso{\VA}^{-1}}{x}}}}}
\\
\Ciso{\CI} \: & = \: \llam{z}{\CI}{z} 
\\
\Ciso{\Cbang{\VA}} \: & = \: \llam{z}{\Ccopower{(\CpsVVT{\VA})}{\CI}}
      {\, \copowerlet{x}{y}{z}{\, \Ilet{y}{\, \bang{(\Vappl{\Viso{\VA}}{x})}}}}
\\
\Ciso{\Ccopower{\VA}{\CB}} \: & = \: \llam{z}{\Ccopower{(\CpsVVT{\VA})}{\CpsCCT{\CB}}}
      {\, \copowerlet{x}{y}{z}{\, \copowerterm{(\Vappl{\Viso{\VA}}{x})}{(\lappl{\Ciso{\CB}\,}{y})}}}
\\
\Ciso{\Czero} \: & = \: \llam{z}{\Czero}{z} 
\\
\Ciso{\CA \Cplus \CB} \: & = \: \llam{z}{\CpsCCT{\CA} \Cplus \CpsCCT{\CB}}
      {\, \Ccase{z}{x}{\,\Cinl{\lappl{\Ciso{\CA}\,}{x}}}{y}{\,\Cinr{\lappl{\Ciso{\CB}\,}{y}}}}
\end{align*}
\caption{Type isomorphisms for the involution property}
\label{fig:tiip}
\end{figure}

\begin{lem}
Suppose $\CR$ is either a computation-type constant or $\CI$. Then
each term $\Viso{\VA} \colon \CpsVVT{\VA} \Vfun \VA$ is an isomorphism, and each
$\Ciso{\CA} \colon \CpsCCT{\CA} \lpop \CA$
is a linear isomorphism.
\end{lem}

\proof
The two statements are proved simultaneously by induction on the type,
with, in the case of a computation type $\CA$, the inverse
for $\Ciso{\CA}$ being established before that of $\Viso{\CA}$.
The assumption that $\CR$ is either a computation-type constant or $\CI$ implies
that $\CpsCT{\CR} = \CI$, and this fact is used frequently
in the proof. We consider just two illustrative cases:
$\Viso{\CA}$ and $\Ciso{\,\Cbang{\VA}}$.

In the case of  $\Viso{\CA}$, we have $\CpsVVT{\CA} = \CpsVT{(\CpsCT{\CA} \lpop \CR)}
= \CpsCT{\CR} \lpop \CpsCCT{\CA} = \CI \lpop \CpsCCT{\CA}$,
and the inverse $\Viso{\CA}^{-1} \colon \CA \Vfun (\CI \lpop \CpsCCT{\CA})$ is defined 
by
\[
\Viso{\CA}^{-1} \: = \: \Vlam{x}{\CA}{\,\llam{z}{\CI}{\, \Ilet{z}{\lappl{\Ciso{\CA}^{-1}}{x}}}} \enspace .
\]
Then we have (using the obvious definition for composition):
\begin{align*}
\Viso{\CA}^{-1} \circ \Viso{\CA} \:
& = \:
\Vlam{h}{\CI \lpop \CpsCCT{\CA}}{\,\llam{z}{\CI}{\, \Ilet{z}{\lappl{\Ciso{\CA}^{-1}}{\lappl{\Ciso{\CA}\,}{\lappl{h\,}{\Itop}}}}}} 
\\
& = \:
\Vlam{h}{\CI \lpop \CpsCCT{\CA}}{\,\llam{z}{\CI}{\, \Ilet{z}{\lappl{h\,}{\Itop}}}} 
& & \text{by induction hypothesis}
\\
& = \:
\Vlam{h}{\CI \lpop \CpsCCT{\CA}}{\,\llam{z}{\CI}{\, \lappl{h\,}{z}}} 
\\
& = \:
\Vlam{h}{\CI \lpop \CpsCCT{\CA}}{\,h} \enspace ,
\end{align*}
and the verification that 
$\Viso{\CA} \circ \Viso{\CA}^{-1} = \Vlam{x}{\!\CA}{x}$ is similarly straightforward.

In the case of  $\Ciso{\,\Cbang{\VA}}$, we have 
$\CpsCCT{(\Cbang{\VA})} = \Ccopower{(\CpsVVT{\VA})}{\CI}$
and the inverse $\Ciso{\,\Cbang{\VA}}^{-1} \colon \Cbang{\VA} \lpop \Ccopower{(\CpsVVT{\VA})}{\CI}$
is defined by
\begin{equation}
\label{banginverse}
\Ciso{\,\Cbang{\VA}}^{-1} \: = \: 
 \llam{w}{\Cbang{\VA}}{\,\banglet{x}{w}{\copowerterm{(\Vappl{\Viso{\VA}^{-1}}{x})}{\Itop}}} \enspace .
\end{equation}
Then:
\begin{align*}
\Ciso{\,\Cbang{\VA}}^{-1} & \circ \Ciso{\,\Cbang{\VA}}
\\
\displaybreak[0]
& = \:\llam{z}{\Ccopower{(\CpsVVT{\VA})}{\CI}}
    {\,\banglet{x'}{(\copowerlet{x}{y}{z}{\Ilet{y}{\bang{(\Vappl{\Viso{\VA}}{x})}}})}{\copowerterm{(\Vappl{\Viso{\VA}^{-1}}{x'})}{\Itop}}}
\\
\displaybreak[0]
& = \:\llam{z}{\Ccopower{(\CpsVVT{\VA})}{\CI}}
    {\,\copowerlet{x}{y}{z}{\Ilet{y}{\banglet{x'}{\bang{(\Vappl{\Viso{\VA}}{x})}}
       {\copowerterm{(\Vappl{\Viso{\VA}^{-1}}{x'})}{\Itop}}}}}
\\
\displaybreak[0]
& = \:\llam{z}{\Ccopower{(\CpsVVT{\VA})}{\CI}}
    {\,\copowerlet{x}{y}{z}{\Ilet{y}{\copowerterm{(\Vappl{\Viso{\VA}^{-1}}{\Vappl{\Viso{\VA}}{x}})}{\Itop}}}}
\\
\displaybreak[0]
& = \:\llam{z}{\Ccopower{(\CpsVVT{\VA})}{\CI}}
    {\,\copowerlet{x}{y}{z}{\Ilet{y}{\copowerterm{x}{\Itop}}}}
\\
\displaybreak[0]
& = \:\llam{z}{\Ccopower{(\CpsVVT{\VA})}{\CI}}
    {\,\copowerlet{x}{y}{z}{\copowerterm{x}{y}}}
\\
\displaybreak[0]
& = \:\llam{z}{\Ccopower{(\CpsVVT{\VA})}{\CI}}
    {\,z}
 \enspace ,
\end{align*}
where the third equality applies the induction hypothesis, and all others,
including the rearrangement of ``let'' expressions in the second equation, justified 
by the equalities of 
Figure~\ref{figure:effects:equalities}.
The verification that 
$\Ciso{\,\Cbang{\VA}} \circ \Ciso{\,\Cbang{\VA}}^{-1}  = \Vlam{x}{\Cbang{\VA}}{x}$ is 
straightforward.
\qed

We remark that the main reason for including  $\CI$ as a primitive EEC construct, in the present paper,
was to permit the  uniform definition of the type isomorphisms, given in Figure~\ref{fig:tiip},
which covers both  cases of interest: when $\CR$ is a computation-type constant, and when it is
$\CI$. The alternative would have beeen to have omitted $\CI$ from the primitive syntax, 
defining it as $\Cbang{\Vone}$. Had this been done, 
we would have  obtained: $\CpsCCT{\CR} = \Vone \!\Cfun \!\CR$, in the case that $\CR$ is
a computation-type constant; and $\CpsCCT{\CR} = {\Ccopower{\Vone}{(\Vone \! \Cfun \! \CR})}$, in the case that
$\CR$ is $\CI$ (i.e., $\CR = \Cbang{\Vone}$). In both cases, 
linear isomorphisms between $\CR$ and $\CpsCCT{\CR}$ still exist, 
they can no longer be given uniformly.

In order to state the fundamental involution property enjoyed
by the self-translation on EEC, for a context
\[
\Gamma \: = \: \In{x_1}{\VC_1}, \dots, \In{x_n}{\VC_n} \enspace ,
\]
we introduce the notation $[\Vappl{\Viso{}^{-1}}{\Gamma}]$ for the substitution
\[
[ \, \Vappl{\Viso{\VC_1}^{-1}}{x_1}, \dots, \Vappl{\Viso{\VC_n}^{-1}}{x_n} \, / \, x_1 , \dots , x_n\,]
\enspace .
\]

\begin{thm}[Involution property]
\label{theorem:involution}
Suppose $\CR$ is either a computation-type constant or $\CI$.
\begin{enumerate}[\em(1)]
\item \label{inv:i}
\label{item:Vnat} If $\Vj{\Gamma}{t}{\VA}$ then
$\Veq{\Gamma}{t}{\Vappl{\Viso{\VA}}{\CpsVVT{t}}\, [\Vappl{\Viso{}^{-1}}{\Gamma}]}{\VA}$.

\item \label{inv:ii} %\label{item:Cnat} 
If $\Cj{\Gamma}{\In{z}{\CA}}{t}{\CB}$ then
$\Ceq{\Gamma}{\In{z}{\CA}}{t}
   {\lappl{\Ciso{\CB}\,}{\,\CpsCCT{t}} \, [\lappl{\Ciso{\CA}^{-1}}{z} \, / \, k_{k_z}]\, [\Vappl{\Viso{}^{-1}}{\Gamma}]\,}
     {\CB}$.
\end{enumerate}
\end{thm}
\proof
The statements are proved simultaneously, by induction on $t$. There are 41 cases in the
proof, one for each of the equations in Figures~\ref{fig:cps:compterms} 
and~\ref{fig:cps:valterms}. By way of illustration,
we verify two of them, the second being among the most complex cases in the proof.

For the first case, suppose $\Vj{\Gamma}{t}{\VA}$. We verify that:
\[\Veq{\Gamma}{\bang{t}}{\Vappl{\Viso{\Cbang{\VA}}}{\CpsVVT{(\bang{t})}}\, [\Vappl{\Viso{}^{-1}}{\Gamma}]}{\bang{\VA}} \enspace .\]
The basic strategy is to first expand the inner $\CpsVT{(\cdot)}$, then the
outer $\CpsVT{(\cdot)}$, applying the 
definitions of $\Viso{\Cbang{\VA}}$ and
$\Ciso{\Cbang{\VA}}$ until the induction hypothesis can be invoked.
Between these steps, we use 
the equalities of Figure~\ref{figure:effects:equalities} to simplify the terms as far as possible.
Henceforth, we treat applications of equalities from
Figure~\ref{figure:effects:equalities} as trivial. So, in the detailed
derivation below, we do not annotate such steps. Nor do we explain 
obvious expansions of $\CpsVT{(\cdot)}$ and $\CpsCT{(\cdot)}$.
\begin{align*}
\Vappl{\Viso{\Cbang{\VA}} &}{\CpsVVT{(\bang{t})}}\, [\Vappl{\Viso{}^{-1}}{\Gamma}] 
\\
\displaybreak[0]
& = \: \Vappl{\Viso{\Cbang{\VA}}}{\CpsVT{(\llam{k}{\CpsVT{\VA} \Cfun \CR}
   {\,\Cappl{k\,}{\CpsVT{t}}})}}\, [\Vappl{\Viso{}^{-1}}{\Gamma}]
\\
\displaybreak[0]
& = \: \Vappl{\Viso{\Cbang{\VA}}}{
     \llam{k'}{\CpsCT{\CR}}{\,\CpsCT{(\Cappl{k\,}{\CpsVT{t}})}\,[k'/k_k]}}\, [\Vappl{\Viso{}^{-1}}{\Gamma}]
\\
\displaybreak[0]
& = \: \Vappl{\Viso{\Cbang{\VA}}}{
     \llam{k'}{\CpsCT{\CR}}{\,\CpsCT{k} \, [\copowerterm{(\CpsVVT{t})}{k_k}\, / k_k]\, \,[k'/k_k]}}\, [\Vappl{\Viso{}^{-1}}{\Gamma}]
\\
\displaybreak[0]
& = \: \Vappl{\Viso{\Cbang{\VA}}}{
     \llam{k'}{\CpsCT{\CR}}{\,k_k \, [\copowerterm{(\CpsVVT{t})}{k_k}\, / k_k]\, \,[k'/k_k]}}\, [\Vappl{\Viso{}^{-1}}{\Gamma}]
\\
\displaybreak[0]
& = \: \Vappl{\Viso{\Cbang{\VA}}}{
     \llam{k'}{\CpsCT{\CR}}{\,\copowerterm{(\CpsVVT{t})}{k'}}} \, [\Vappl{\Viso{}^{-1}}{\Gamma}]
\\
\displaybreak[0]
& = \: \lappl{\Ciso{\Cbang{\VA}}}{
     \lappl{(\llam{k'}{\CpsCT{\CR}}{\,\copowerterm{(\CpsVVT{t})}{k'}) \,}}
         {\Itop}} \, [\Vappl{\Viso{}^{-1}}{\Gamma}]
& & \text{def.~of $\Viso{\Cbang{\VA}}$}
\\
\displaybreak[0]
& = \: \lappl{\Ciso{\Cbang{\VA}}}{\copowerterm{(\CpsVVT{t})}{\Itop}} \, [\Vappl{\Viso{}^{-1}}{\Gamma}]
\\
\displaybreak[0]
& = \: \copowerlet{x}{y}{\copowerterm{(\CpsVVT{t})}{\Itop}}{\, \Ilet{y}{\, \bang{(\Vappl{\Viso{\VA}}{x})}}}
\, [\Vappl{\Viso{}^{-1}}{\Gamma}]
& & \text{def.~of $\Ciso{\Cbang{\VA}}$}
\\
\displaybreak[0]
& = \: \Ilet{\Itop}{\, \bang{(\Vappl{\Viso{\VA}}{\CpsVVT{t}})}}\, [\Vappl{\Viso{}^{-1}}{\Gamma}]
\\
\displaybreak[0]
& = \:  \bang{(\Vappl{\Viso{\VA}}{\CpsVVT{t}})}\, [\Vappl{\Viso{}^{-1}}{\Gamma}]
\\
\displaybreak[0]
& = \:  \bang{(\Vappl{\Viso{\VA}}{\CpsVVT{t}}\, [\Vappl{\Viso{}^{-1}}{\Gamma}])}
\\
\displaybreak[0]
& = \: \bang{t} 
& & \text{induction hypothesis}
\enspace .
\end{align*}

For the second case, suppose $\Cj{\Gamma}{\In{z}{\CD}}{t}{\Cbang{\VA}}$ and 
$\Vj{\Gamma, \In{x}{\VA}}{u}{\CB}$. We verify that
\[
\Ceq{\Gamma}{\In{z}{\CD}}{\banglet{x}{t}{u}\,}
    {\, \lappl{\Ciso{\CB}\,}{\CpsCCT{(\banglet{x}{t}{u})}}
         \, [\lappl{\Ciso{\CB}^{-1}}{z} \, / \, k_{k_z}]\, [\Vappl{\Viso{}^{-1}}{\Gamma}]\, } {\CB}
\enspace .
\]
Adopting a similar strategy to above, we obtain:
\begin{align}
\nonumber
\lappl{\Ciso{\CB} & \,}{\CpsCCT{(\banglet{x}{t}{u})}}
         \, [\lappl{\Ciso{\CD}^{-1}}{z}  /  k_{k_z}]\, [\Vappl{\Viso{}^{-1}}{\Gamma}]
\\
\displaybreak[0]
\nonumber
& = \: 
\lappl{\Ciso{\CB}\,}{\CpsCT{(\CpsCT{t}\,[(\Clam{x}{\CpsVT{\VA}}{\,\lappl{\CpsVT{u}\,}{k_z}})\, / \, k_z])}}
         \, [\lappl{\Ciso{\CD}^{-1}}{z}  /  k_{k_z}]\, [\Vappl{\Viso{}^{-1}}{\Gamma}]
\\
\displaybreak[0]
\label{complicated:i}
& = \: 
\lappl{\Ciso{\CB}\,}
   {\CpsCT{(\Clam{x}{\CpsVT{\VA}}{\,\lappl{\CpsVT{u}\,}{k_z}})} \,[\CpsCCT{t}\, / \, k_{k_z}]}
         \, [\lappl{\Ciso{\CD}^{-1}}{z}  /  k_{k_z}]\, [\Vappl{\Viso{}^{-1}}{\Gamma}]
\\
\displaybreak[0]
\nonumber
& = \: 
\lappl{\Ciso{\CB}\,}
   {(\copowerlet{x}{h}{k_{k_z}}{\CpsCT{(\lappl{\CpsVT{u}\,}{k_z})}\, [h/{k_{k_z}}]}) \,[\CpsCCT{t}\, / \, k_{k_z}]}
         \, [\lappl{\Ciso{\CD}^{-1}}{z}  /  k_{k_z}]\, [\Vappl{\Viso{}^{-1}}{\Gamma}]
\\
\displaybreak[0]
\nonumber
& = \: 
\lappl{\Ciso{\CB}\,}
   {(\copowerlet{x}{h}{\CpsCCT{t}}{\CpsCT{(\lappl{\CpsVT{u}\,}{k_z})}\, [h/{k_{k_z}}]})}
         \, [\lappl{\Ciso{\CB}^{-1}}{z}  /  k_{k_z}]\, [\Vappl{\Viso{}^{-1}}{\Gamma}]
\\
\displaybreak[0]
\nonumber
& = \: 
\lappl{\Ciso{\CB}\,}
   {(\copowerlet{x}{h}{\CpsCCT{t}}{(\CpsCT{k_z} \, [\lappl{\CpsVVT{u}\,}{k_{k_z}} \, / \, k_{k_z}])\, [h/{k_{k_z}}]})}
         \, [\lappl{\Ciso{\CD}^{-1}}{z}  /  k_{k_z}]\, [\Vappl{\Viso{}^{-1}}{\Gamma}]
\\
\displaybreak[0]
\nonumber
& = \: 
\lappl{\Ciso{\CB}\,}
   {(\copowerlet{x}{h}{\CpsCCT{t}}{(k_{k_z} \, [\lappl{\CpsVVT{u}\,}{k_{k_z}} \, / \, k_{k_z}])\, [h/{k_{k_z}}]})}
         \, [\lappl{\Ciso{\CD}^{-1}}{z}  /  k_{k_z}]\, [\Vappl{\Viso{}^{-1}}{\Gamma}]
\\
\displaybreak[0]
\nonumber
& = \: 
\lappl{\Ciso{\CB}\,}
   {(\copowerlet{x}{h}{\CpsCCT{t}}{\lappl{\CpsVVT{u}\,}{k_{k_z}}\, [h/{k_{k_z}}]})}
         \, [\lappl{\Ciso{\CD}^{-1}}{z}  /  k_{k_z}]\, [\Vappl{\Viso{}^{-1}}{\Gamma}]
\\
\displaybreak[0]
\nonumber
& = \: 
\lappl{\Ciso{\CB}\,}
   {(\copowerlet{x}{h}{\CpsCCT{t}}{\lappl{\CpsVVT{u}\,}{h}})}
         \, [\lappl{\Ciso{\CD}^{-1}}{z}  /  k_{k_z}]\, [\Vappl{\Viso{}^{-1}}{\Gamma}]
\\
\displaybreak[0]
\nonumber
& = \: 
   (\copowerlet{x}{h}{\CpsCCT{t}}{\lappl{\Ciso{\CB}\,}{\lappl{\CpsVVT{u}\,}{h}}})
         \, [\lappl{\Ciso{\CD}^{-1}}{z}  /  k_{k_z}]\, [\Vappl{\Viso{}^{-1}}{\Gamma}]
\\
\displaybreak[0]
\nonumber
& = \: 
   (\copowerlet{x}{h}{\CpsCCT{t}}
      {\Ilet{h}{\lappl{\Ciso{\CB}\,}{\lappl{\CpsVVT{u}\,}{\Itop}}}})
         \, [\lappl{\Ciso{\CD}^{-1}}{z}  /  k_{k_z}]\, [\Vappl{\Viso{}^{-1}}{\Gamma}]
\\
\displaybreak[0]
\label{complicated:ii}
& = \: 
   (\copowerlet{x}{h}{\CpsCCT{t}}
      {\Ilet{h}{\lappl{\Viso{\CB}\,}{\CpsVVT{u}}}})
         \, [\lappl{\Ciso{\CD}^{-1}}{z}  /  k_{k_z}]\, [\Vappl{\Viso{}^{-1}}{\Gamma}]
\\
\displaybreak[0]
\nonumber
& = \: 
   \copowerlet{x}{h}{(\CpsCCT{t}\, [\lappl{\Ciso{\CD}^{-1}}{z}  /  k_{k_z}]\, [\Vappl{\Viso{}^{-1}}{\Gamma}])}
     {\Ilet{h}{(\lappl{\Viso{\CB}\,}{\CpsVVT{u}}\, [\Vappl{\Viso{}^{-1}}{\Gamma}])}}
\\
\displaybreak[0]
\label{complicated:iii}
& = \: 
   \copowerlet{x}{h}{\lappl{\Ciso{\,\Cbang{\VA}}^{-1}}{t}}{\Ilet{h}{(\lappl{\Viso{\CB}\,}{\CpsVVT{u}}\, [\Vappl{\Viso{}^{-1}}{\Gamma}])}}
\\
\displaybreak[0]
\label{complicated:iv}
& = \: 
   \copowerlet{x}{h}{(\banglet{x}{t}{\copowerterm{(\Vappl{\Viso{\VA}^{-1}}{x})}{\Itop}})}{\Ilet{h}{(\lappl{\Viso{\CB}\,}{\CpsVVT{u}}\, [\Vappl{\Viso{}^{-1}}{\Gamma}])}}
\\
\displaybreak[0]
\nonumber
& = \: 
   \banglet{x}{t}
     {\copowerlet{x}{h}{\copowerterm{(\Vappl{\Viso{\VA}^{-1}}{x})}{\Itop}}
         {\Ilet{h}{(\lappl{\Viso{\CB}\,}{\CpsVVT{u}}\, [\Vappl{\Viso{}^{-1}}{\Gamma}])}}}
\\
\displaybreak[0]
\nonumber
& = \: 
   \banglet{x}{t}
         {\Ilet{\Itop}{(\lappl{\Viso{\CB}\,}{\CpsVVT{u}}
             \, [\Vappl{\Viso{}^{-1}}{\Gamma}] \,[\Vappl{\Viso{\VA}^{-1}}{x} \, / \, x])}}
\\
\displaybreak[0]
\nonumber
& = \:
   \banglet{x}{t}
         {(\lappl{\Viso{\CB}\,}{\CpsVVT{u}}
             \, [\Vappl{\Viso{}^{-1}}{\Gamma, \In{x}{\VA}}])}
\\
\displaybreak[0]
\label{complicated:v}
& = \:
   \banglet{x}{t}{u}
\enspace .
\end{align}
Here, (\ref{complicated:i}) is by Proposition~\ref{prop:trans:subs}.\ref{subs:iv},
 (\ref{complicated:ii}) is by definition of $\Ciso{\CB}$,
 (\ref{complicated:iii}) applies the induction hypothesis for $t$,
 (\ref{complicated:iv}) expands $\Ciso{\,\Cbang{\VA}}^{-1}$ 
using (\ref{banginverse}), and 
 (\ref{complicated:v}) applies the induction hypothesis for $t$ (which is applicable only 
at this point in the argument, because $t$ is typed 
relative to the context $\Gamma,\In{x}{\VA}$ rather than $\Gamma$).
\qed

We end the present section by applying Theorem~\ref{theorem:involution}
to derive the full completeness of
the self-translation.

\begin{thm}[Full completeness of self-translation] 
\label{thm:full:complete}
Suppose $\CR$ is either a computation-type constant or $\CI$.
\begin{enumerate}[\em(1)]
\item \label{fc:i}
  If $\Vj{\Gamma}{t,u}{\VA}$ and 
$\Veq{\CpsVT{\Gamma}}{\CpsVT{t}}{\CpsVT{u}}{\CpsVT{\VA}}$ 
then $\Veq{\Gamma}{t}{u}{\VA}$.

\item \label{fc:ii}
       If $\Vj{\CpsVT{\Gamma}}{t}{\CpsVT{\VA}}$ 
       then there exists  $\Vj{\Gamma}{u}{\VA}$ such that 
$\Veq{\CpsVT{\Gamma}}{t}{\CpsVT{u}}{\CpsVT{\VA}}$.

\item \label{fc:iii}
If $\Cj{\Gamma}{\In{z}{\CA}}{t,u}{\CB}$ and  
$\Ceq{\CpsVT{\Gamma}}{\In{k_z}{\CpsCT{\CB}}}{\CpsCT{t}}{\CpsCT{u}}{\CpsCT{\CA}}$
then $\Ceq{\Gamma}{\In{z}{\CA}}{t}{u}{\CB}$.

\item \label{fc:iv}
If $\Cj{\CpsVT{\Gamma}}{\In{k_z}{\CpsCT{\CB}}}{t}{\CpsCT{\CA}}$ then there exists 
$\Cj{\Gamma}{\In{z}{\CA}}{u}{\CB}$ such that \\
$\Ceq{\CpsVT{\Gamma}}{\In{k_z}{\CpsCT{\CB}}}{t}{\CpsCT{u}}{\CpsCT{\CA}}$.
\end{enumerate}
\end{thm}
\proof
For statement~\ref{fc:i}, suppose $\Vj{\Gamma}{t,u}{\VA}$ and $\CpsVT{t} = \CpsVT{u}\,$. Then:
\begin{align*}
t & = \Vappl{\Viso{\VA}\,}{\CpsVVT{t}\, [\Vappl{\Viso{}^{-1}}{\Gamma}]}
 && \text{(Theorem~\ref{theorem:involution}.\ref{inv:i})} 
\\
& = \Vappl{\Viso{\VA}\,}{\CpsVVT{u}\, [\Vappl{\Viso{}^{-1}}{\Gamma}]}
 && \text{(Theorem~~\ref{thm:cps:soundness})} 
\\
 & = u 
 && \text{(Theorem~\ref{theorem:involution}.\ref{inv:i})} \enspace .
\end{align*}
\noindent
For statement~\ref{fc:ii}, suppose $\Vj{\CpsVT{\Gamma}}{t}{\CpsVT{\VA}}$.
Define $u = \Vappl{\Viso{\VA}}{\CpsVT{t}\,[\Vappl{\Viso{}^{-1}}{\Gamma}]}$. Then:
\begin{align*}
\CpsVT{u} & =  \Vappl{\Viso{{\CpsVT{\VA\!}}}}{\CpsVVVT{u} \, [\Vappl{\Viso{}^{-1}}{\CpsVT{\Gamma}}]}
 && \text{(Theorem~\ref{theorem:involution}.\ref{inv:i})} \\
& = \Vappl{\Viso{{\CpsVT{\VA\!}}}}
      {\CpsVT{( \Vappl{\Viso{\CA}^{-1}}{u \, [\Viso{}(\Gamma)]})} \,[\Vappl{\Viso{}^{-1}}{\CpsVT{\Gamma}}]}
 && \text{(Theorem~\ref{theorem:involution}.\ref{inv:i})} \\
& = \Vappl{\Viso{{\CpsVT{\VA\!}}}}{\CpsVVT{t} \, [\Vappl{\Viso{}^{-1}}{\CpsVT{\Gamma}}]}
 && \text{(Definition of $u$)} \\
& = t 
 && \text{(Theorem~\ref{theorem:involution}.\ref{inv:i})} \enspace .
\end{align*}
The proofs of statements~\ref{fc:iii} and~\ref{fc:iv} are similar.
\qed

\def\Groupoid{\mathbf{Grpd}}
\def\cni{{cni}}
\def\fudgey{sufficiently bicomplete}

\section{Recovering linear-use CPS translations of typed lambda-calculus}
\label{section:recovering}

In this section, we use the self-translation to establish properties
of the call-by-value and call-by-name linear-use CPS translations
of Section~\ref{sec:lin:cps}. The main property we exploit 
is that  the generic self-translation 
subsumes  the call-by-value
and call-by-name translations.
Indeed, the latter are obtained uniformly by precomposing the
generic self-translation on EEC with 
the standard call-by-value and call-by-name translations
from $\lambda$-calculus to EEC, given in Section \ref{section:cbv:cbn}.

\begin{thm}[Recovering $\cbvLincps{(\cdot)}$]
\label{theorem:recover:cbv}
For every simple type $\sigma$, we have 
$\cbvLincps{\sigma} \, = \, \CpsVTcbv{\sigma}$; and, for
every simply-typed term $\Lj{\Theta}{M}{\sigma}$, we have 
$\Veq{\cbvLincps{\Theta}}{\cbvLincps{M}}{\CpsVTcbv{M}}{(\cbvLincps{\sigma} \Cfun \CR) \lpop \CR}$.
\end{thm}

\begin{thm}[Recovering $\cbnLincps{(\cdot)}$]
\label{theorem:recover:cbn}
Suppose $\CR$ is different from $\comptype{\alpha}$, for every
simply-typed $\lambda$-calculus type constant $\alpha$.
Then, for every $\sigma$, 
we have 
$\cbnLincps{\sigma} \, = \, \CpsCTcbn{\sigma}$, hence
$\cbnLincps{\sigma} \lpop \CR \, = \, \CpsVTcbn{\sigma}$; and, for
every term $\Lj{\Theta}{M}{\sigma}$, we have 
$\Veq{\cbnLincps{\Theta} \lpop \CR}{\cbnLincps{M}}{\CpsVTcbn{M}}{\cbnLincps{\sigma}\lpop \CR}$.
\end{thm}
\noindent
The  proofs are by induction on the structure of $\sigma$ and $M$. 
\proof[Proof of Theorem~\ref{theorem:recover:cbv}]
For the type equality,  we consider the case of $\sigma \Lfun \tau$.
\begin{align*}
\CpsVTcbv{(\sigma \Lfun \tau)} \: & = \: \CpsVT{(\cbv{\sigma} \Vfun \bang{\,\cbv{\tau}})}
\\
& = \: \CpsVTcbv{\sigma} \Vfun \CpsVT{(\bang{\,\cbv{\tau}})} 
\\
& = \: \CpsVTcbv{\sigma} \Vfun (\CpsCT{(\bang{\,\cbv{\tau}})}  \lpop \CR)
\\
& = \: \CpsVTcbv{\sigma} \Vfun ((\CpsVTcbv{\tau} \Cfun \CR)  \lpop \CR)
\\ 
& = \: \cbvLincps{\sigma} \Vfun ((\cbvLincps{\tau} \Cfun \CR)  \lpop \CR)
& & \text{by induction hypothesis}
\\
& = \: \cbvLincps{(\sigma \Lfun \tau)} \enspace .
\end{align*}
And, in the case of the term $\Lappl{M}{N}$, where $\Lj{\Theta}{M}{\sigma \Lfun \tau}$
and $\Lj{\Theta}{N}{\sigma}$, we have:
\begin{align*}
\CpsVTcbv{(\Lappl{M}{N})} \: 
& = \:  \CpsVT{(\banglet{f}{\cbv{M}}{\banglet{x}{\cbv{N}}{\Vappl{f}{x}}})}
\\
& = \: \llam{k}{\CpsVTcbv{\tau} \Cfun \CR}
   {\,\lappl{\CpsVTcbv{M}\,}{\Clam{f}{\CpsVTcbv{(\sigma \Lfun \tau)}}
       {\,\lappl{\CpsVT{(\banglet{x}{\cbv{N}}{\Vappl{f}{x}})}\,}{k}}}}
\\
& = \: \llam{k}{\CpsVTcbv{\tau} \Cfun \CR}
   {\,\lappl{\CpsVTcbv{M}\,}{\Clam{f}{\CpsVTcbv{(\sigma \Lfun \tau)}}
       {\,\lappl{\CpsVTcbv{N}\,}{\Clam{x}{\CpsVTcbv{\sigma}}{\,\lappl{\Vappl{f}{x}}{k}}}}}}
\\
& = \: \llam{k}{\cbvLincps{\tau} \Cfun \CR}
     {\,\lappl{\cbvLincps{M}\,}{\Clam{f}{\cbvLincps{\sigma} \Vfun (\cbvLincps{\tau} \Cfun \CR) \lpop \CR}
         {\,\lappl{\cbvLincps{N}\,}{\Clam{x}{\cbvLincps{\sigma}}{\,\lappl{\Vappl{f}{x}}{k}}}}}} 
\\
& = \: \cbvLincps{(\Lappl{M}{N})} \enspace .
\end{align*}
\qed
\proof[Proof of Theorem~\ref{theorem:recover:cbn}]
First, we observe that for a type-constant $\alpha$, we have
\[\cbnLincps{\alpha} = \comptype{\alpha} =  \CpsCT{\comptype{\alpha}} = \CpsCTcbn{\alpha} \enspace , \]
where the middle equality relies on the asumption that 
$\comptype{\alpha}$ is different from $\CR$.

Of the other cases, we again consider $\sigma \Lfun \tau$.
\begin{align*}
\CpsCTcbn{(\sigma \Lfun \tau)} \: & = \: \CpsCT{(\cbn{\sigma} \Cfun \cbn{\tau})}
\\
& = \: \Ccopower{(\CpsVTcbn{\sigma})}{\CpsCTcbn{\tau}}
\\
& = \: \Ccopower{(\CpsCTcbn{\sigma} \lpop \CR)}{\CpsCTcbn{\tau}}
\\ & = \: \Ccopower{(\cbnLincps{\sigma} \lpop \CR)}{\cbnLincps{\tau}}
& & \text{by induction hypothesis}
\\
& = \: \cbnLincps{(\sigma \Lfun \tau)} \enspace .
\end{align*}
And the case of an application $\Lappl{M}{N}$ works out as:
\begin{align*}
\CpsVTcbn{(\Lappl{M}{N})} \: 
& = \:  \CpsVT{(\Cappl{\cbn{M}\,}{\cbn{N}})}
\\
& = \: \llam{k}{\CpsCTcbn{\tau}}
   {\,\lappl{\CpsVTcbn{M}\,}{\copowerterm{(\CpsVTcbn{N})\,}{\,k}}}
\\
& = \: \llam{k}{\cbnLincps{\tau}}
   {\,\lappl{\cbnLincps{M}\,}{\copowerterm{(\cbnLincps{N})\,}{\,k}}}
\\
& = \: \cbnLincps{(\Lappl{M}{N})} \enspace .
\end{align*}
\qed

We comment that many of the syntactic choices of this paper have been  made 
in order to obtain Theorems~\ref{theorem:recover:cbv} and~\ref{theorem:recover:cbn}
in the simple form stated. For example, in the conference version of the paper~\cite{EMS:fossacs},
where neither value-type and computation-type products nor
value-\ and computation-type function spaces are distinguished syntactically,
Theorem~\ref{theorem:recover:cbn} holds only up to type isomorphism,
rather than up to equality. Similarly, had a different choice been made for
$\cbvLincps{(\sigma \Lfun \tau)}$
in Figure~\ref{figure:lincps}, for example
$(\cbvLincps{\tau} \Cfun \CR) \lpop  (\cbvLincps{\sigma} \Cfun  \CR)$,
as discussed in Section~\ref{sec:lin:cps}, then Theorem~\ref{theorem:recover:cbv}
would have held only up to isomorphism. 

Using Theorems~\ref{theorem:recover:cbv} and~\ref{theorem:recover:cbn}, it is now straightforward
to provide the postponed proofs of soundness
and full completeness for the cbv and cbn linear-use CPS translations of simply-typed $\lambda$-calculus from
Section~\ref{sec:lin:cps}, by deriving  these
results as consequences of soundness and full completeness for the self-translation
(Theorems~\ref{thm:cps:soundness} and~\ref{thm:full:complete}).
We give the proofs for the call-by-value case only 
(Proposition~\ref{prop:cbvLincps:sound} and Theorem~\ref{thm:cbvLincps:complete}).
The proofs for the corresponding call-by-name results 
(Proposition~\ref{prop:cbnLincps:sound} and Theorem~\ref{thm:cbnLincps:complete}),
are similarly straightforward.

\proof[Proof of Proposition~\ref{prop:cbvLincps:sound} (Soundness $\cbvLincps{(\cdot)}$)]
Suppose
$\LCeq{\Theta}{M}{N}{\tau}$. Proposition~\ref{prop:cbv}.\ref{cbv:i} shows
$\Veq{\cbv{\Theta}}{\cbv{M}}{\cbv{N}}{\Cbang{(\cbv{\tau})}}$.
Whence, by Theorem~\ref{thm:cps:soundness},
$\Veq{\CpsVT{(\cbv{\Theta})}}{\CpsVT{(\cbv{M})}}{\CpsVT{(\cbv{N})}}{\CpsVT{(\Cbang{\,\cbv{\tau}})}}$.
That is, by Theorem~\ref{theorem:recover:cbv},
$\Veq{\cbvLincps{\Theta}}{\cbvLincps{M}}{\cbvLincps{N}}{(\cbvLincps{\tau} \Cfun \CR) \lpop \CR}$.
\qed

\proof[Proof of Theorem~\ref{thm:cbvLincps:complete}  (Full completeness of $\cbvLincps{(\cdot)}$)]
For statement~\ref{fcl:i}, suppose $\Lj{\Theta}{M,N}{\tau}$ and 
$\Veq{\cbvLincps{\Theta}}{\cbvLincps{M}}{\cbvLincps{N}}{(\cbvLincps{\tau} \Cfun \CR) \lpop \CR}$.
By Theorem~\ref{theorem:recover:cbv}, this is equivalent to
$\Veq{\CpsVTcbv{\Theta}}{\CpsVTcbv{M}}{\CpsVTcbv{N}}{\CpsVT{(\Cbang{\,\cbv{\tau}})}}$.
So, by Theorem~\ref{thm:full:complete}.\ref{fc:i},
$\Veq{\cbv{\Theta}}{\cbv{M}}{\cbv{N}}{\Cbang{(\cbv{\tau})}}$.
Whence, by Proposition~\ref{prop:cbv}.\ref{cbv:ii},
$\LCeq{\Theta}{M}{N}{\tau}$, as required.

For statement~\ref{fcl:ii}, 
suppose 
$\Vj{\cbvLincps{\Theta}}{t}{(\cbvLincps{\tau} \Cfun \CR) \lpop \CR}$.
That is, by Theorem~\ref{theorem:recover:cbv},
$\Vj{\CpsVTcbv{\Theta}}{t}{\CpsVT{(\Cbang{\, \cbv{\tau}})}}$.
Then, by Theorem~\ref{thm:full:complete}.\ref{fc:ii} there exists 
$\Vj{\cbv{\Theta}}{u}{\Cbang{\,\cbv{\tau}}}$ such that
$\Veq{\CpsVTcbv{\Theta}}{t}{\CpsVT{u}}{\CpsVT{(\Cbang{\, \cbv{\tau}})}}$.
And, by Proposition~\ref{prop:cbv}.\ref{cbv:iii},
there exists $\Lj{\Theta}{M}{\tau}$ such that
$\Veq{\cbv{\Theta}}{u}{\cbv{M}}{\Cbang{\,\cbv{\tau}}}$.
Therefore, by Theorem~\ref{thm:cps:soundness},
$\Veq{\CpsVTcbv{\Theta}}{t}{\CpsVTcbv{M}}{\CpsVT{(\Cbang{\, \cbv{\tau}})}}$.
That is, again by Theorem~\ref{theorem:recover:cbv},
$\Veq{\cbvLincps{\Theta}}{t}{\cbvLincps{M}}{(\cbvLincps{\tau} \Cfun \CR) \lpop \CR}$, as required.
\qed

\section{Perspectives}
\label{section:perspectives}

Throughout the paper, we have taken EEC for granted. However, linear-use CPS translations
can themselves be used as a motivation for the selection of type constructors 
appearing in EEC. 
Given Hasegawa's call-by-value and call-by-name linear-use CPS translations into ILL
\cite{Hasegawa:Flops:02,Hasegawa:Flops:04}, it is natural to ask if these translations
can be encompassed within a single linear-use CPS translation of Levy's CBPV
into ILL --- since one of the \emph{raisons d'\^{e}tre} of CBPV is to have a uniform 
language generalising cbv and cbn~\cite{Levy:book}. For our \emph{effect calculus} (EC),
that is, for CBPV without value-type sums (see the discussion in 
Section~\ref{section:calculus}), the answer is provided by
our generic self-translation on EEC. A linear-use CPS translation of the effect calculus is obtained by
restricting the source of the self-translation to EC,
and by reinterpreting the target of the translation as ILL.
Having done this, one sees that
the fragment of ILL that is used in performing this translation is EEC.
Thus EEC arises as naturally the smallest fragment of ILL able to act as a target language
for a linear-use CPS translation of EC.
Value-type sums, that is the whole of CBPV, can be accommodated in the picture
by simply adding value-type sums to EEC, see~\cite{EMSb}. The 
generic self-translation of Section~\ref{section:canonical} easily extends
to a self-translation on the resulting system EEC$+$. Thus there is a
linear-use CPS translation of full CBPV into EEC$+$.\footnote{It 
is less straightforward to give a linear-use
CPS translation of the whole of CBPV into ILL. Because there is
no distinction between ``linear'' and ``intuitionistic'' types, analogous
to the distinction between computation and value types, there is no
natural interpretation for  value-type sums in ILL. Sums are best incorporated by
moving to a version of linear logic that includes such a type distinction~\cite{Benton:95}.}

It is a remarkable fact that EEC supports its own linear-use CPS translation as
a self-translation. As we have seen, this property does not hold of smaller 
fragments, such as the effect calculus, whose linear-use CPS translation
requires the full expressivity of EEC.  It also does not extend to ILL
itself. That is, the linear-use CPS translation of EEC cannot be extended to
obtain an analogous linear-use CPS translation from ILL to itself.
To appreciate this, it is necessary to say something about the
category-theoretic model theory underlying linear-use CPS translations.
This model theory provides an illuminating perspective on the 
syntactic material presented in the paper.

Roughly speaking, a \emph{model} $\mathcal{M}$ of EEC
is given by a tuple 
\[(\mathcal{V}, \, \mathcal{C}, \, F \dashv G \colon \mathcal{C} \to \mathcal{V}) \enspace ,\]
where: $\mathcal{V}$ is a category modelling functions between value types;
$\mathcal{C}$ is a category modelling linear functions between computation types;
and $F \dashv G$ is an adjunction, with $F$  modelling the $\Cbang{(\cdot)}$ type construction,
and $G$ providing the coercion from computation types to value types.
A significant amount of 
additional structure,
all of which is determined by universal properties,
is also required on the categories, to interpret the other type constructors of EEC.
The reader is referred to~\cite{EMS,EMSb,EMSc} for further details,
which are somewhat technical --- substantial use is made of 
{enriched category theory}~\cite{Kelly:book}.
The point relevant to the content of the present paper is that 
the categorical models of EEC are closed under an interesting construction.
Given a model as above, let $\CR$ be a chosen object of $\mathcal{C}$.
We call the structure 
$(\mathcal{V}, \, \mathcal{C},\,  F \dashv G \colon \mathcal{C} \to \mathcal{V}, \, \CR)$ a
\emph{pointed model}. 
Such a pointed model, $\mathcal{N}$, has a \emph{dual (pointed) model}:
\[
\mathcal{N}^{\star} \: = \: 
(\mathcal{V}, \,\mathcal{C}^{\mathrm{op}}, \,
  F^{\star} \dashv G^{\star} \colon \mathcal{C}^{\mathrm{op}} \to \mathcal{V},\, \CI) \enspace ,
\]
where $F^{\star}$ corresponds to the contravariant mapping $\VA \mapsto \VA \Cfun \CR$ from value to computation types,
$G^{\star}$ 
corresponds to the contravariant mapping $\CA \mapsto \CA \lpop \CR$ in the other direction,
and $\CI$ is the object of $\mathcal{C}$ chosen to model the type $\CI$ in $\mathcal{N}$.
%($\CI$ is thus isomorphic to $F\Vone$, where $\Vone$ is the terminal object of $\mathcal{V}$).
Thus the monad $GF$ on $\mathcal{V}$, which models $\Cbang{(\cdot)}$ in $\mathcal{N}$,
is converted to the monad $G^{\star}F^{\star} = ((\cdot) \Cfun \CR) \lpop \CR$ on 
$\mathcal{V}$, which models $\Cbang{(\cdot)}$ in the dual model $\mathcal{N}^{\star}$.
Monads of the form $G^{\star}F^{\star}$ have been called \emph{dual monads} 
by Lawvere~\cite{Lawvere:doctrines}. In our setting, the dual terminology is particularly
apt, since we have:
\begin{fact}
\label{fact:double:dual}
Every pointed model $\mathcal{N}$ is isomorphic to its double dual ${\mathcal{N}^{\star\star}}$.
\end{fact}
\noindent
Importantly, the isomorphism preserves the pointed-model structure, but only up
to coherent natural isomorphism. Up-to-isomorphism structure preservation
is taken as the basic notion of morphism of EEC models~\cite{EMS,EMSc}.
Those  special morphisms that preserve structure up to equality (``on the nose'')
are referred to as \emph{strict}.

Given the description of $G^{\star}F^{\star}$ as $((\cdot) \Cfun \CR) \lpop \CR$,
a connection with linear-use CPS translations is apparent at the level 
of  monads. Accordingly, one might call $\mathcal{N}^{\star}$ a linearly-used continuations
model relative to $\mathcal{N}$. By Fact~\ref{fact:double:dual},
every (pointed) model of EEC arises as a linearly-used continuations model relative
to another model, namely relative to its own dual model --- a property, which is somewhat
surprising at first sight. 

The dual monad construction also allows us to reconstruct the 
self-translation of Section~\ref{section:canonical} semantically.
There is a syntactic model $\mathcal{M}_{\mathrm{syn}}$ 
whose objects are EEC types and whose morphisms are 
terms modulo provable equality. 
This enjoys an initiality property: for any interpretation
of type constants in a model $\mathcal{M}$  there is a unique
strict %\footnote{By strict, we mean all structure is preserved up to equality.}
morphism of models from $\mathcal{M}_{\mathrm{syn}}$ to $\mathcal{M}$
that maps type constants in the specified way. 
Let $\CR$ be a chosen computation type.
Define $\mathcal{N}_{\mathrm{syn}\CR}$ be the pointed model with $\CR$ as 
its point.
Interpret
all type constants as themselves,  except for $\CR$ which, if it is a type constant,
gets interpreted as $\CI$. Then the induced strict morphism of  models
from $\mathcal{M}_{\mathrm{syn}}$ to the underlying model
of $\mathcal{N}_{\mathrm{syn}\CR}^{\star}$ is exactly the generic self-translation of 
Section~\ref{section:canonical}. That is, the action of the morphism 
on (objects and morphisms of) $\mathcal{V}$ is
given by $\CpsVT{(\cdot)}$ (on types and terms respectively), and its action on $\mathcal{C}$
is given by $\CpsCT{(\cdot)}$.

It is now possible to substantiate the claim made earlier that the 
self-translation of EEC does not extend to the whole of ILL.
Any model of ILL (of the general form described in~\cite{Benton:95}) 
is also a model of ECC. Let $\mathcal{N}$ be a pointed model of ILL.
Then, in general, its dual $\mathcal{N}^{\star}$, although still a model of EEC,
is not a model of ILL (the linear category need not be
symmetric monoidal closed). In particular, when $\mathcal{N}$ is the
syntactic (initial) model of ILL (with chosen $\CR$), the dual model
$\mathcal{N}^{\star}$ is not a model of ILL.  Thus there is no induced morphism of models
from the syntactic ILL model to its dual. That is, there is no linear-use CPS
translation of ILL to itself.

Returning to the self-translation of EEC, we now outline how the semantic perspective
provides a conceptually clean proof of the involution property and full completeness.
Suppose $\CR$ is either a type constant or $\CI$. Then the
morphism from $\mathcal{M}_{\mathrm{syn}}$ to $\mathcal{N}_{\mathrm{syn}\CR}^{\star}$,
described above as corresponding to the self-translation, extends (trivially) to a morphism
of pointed models from $\mathcal{N}_{\mathrm{syn}\CR}$ to $\mathcal{N}_{\mathrm{syn}\CR}^{\star}$.
The operation of taking duals is functorial (in an appropriate 2-categorical sense), and so
we obtain a morphism of pointed models from 
$\mathcal{N}_{\mathrm{syn}\CR}^{\star}$ to $\mathcal{N}_{\mathrm{syn}\CR}^{\star\star}$;
whence, by composition, a morphism from 
$\mathcal{N}_{\mathrm{syn}\CR}$ to $\mathcal{N}_{\mathrm{syn}\CR}^{\star\star}$. 
The composite morphism preserves type constants. Furthermore $\mathcal{M}_{\mathrm{syn}}$
enjoys a universal property with respect to non-strict morphisms: 
for any interpretation
of type constants in a model $\mathcal{M}$  there is a 
unique-up-to-coherent-natural-isomorphism 
morphism of models from $\mathcal{M}_{\mathrm{syn}}$ to $\mathcal{M}$
that maps type constants (up to isomorphism) in the specified way. 
This means that, the induced morphism from $\mathcal{N}_{\mathrm{syn}\CR}$ to 
$\mathcal{N}_{\mathrm{syn}\CR}^{\star\star}$ is coherently
naturally isomorphic to the morphism implementing the double-duality of Fact~\ref{fact:double:dual}.
This is literally the involution property of the self-translation (Theorem~\ref{theorem:involution}) in semantic form.
With a little more manipulation of the universal property
of $\mathcal{M}_{\mathrm{syn}}$, one obtains:
\begin{fact}
The morphism 
from $\mathcal{N}_{\mathrm{syn}\CR}$ to $\mathcal{N}_{\mathrm{syn}\CR}^{\star}$ is an equivalence
of pointed models.
\end{fact}

\noindent 
This result corresponds to the full completeness of the self translation
(Theorem~\ref{thm:full:complete}). Semantically, it states the
surprising, at first sight, fact that the syntactic (pointed) model is self-dual.

There is, however,  an alternative perspective on models, from which the
self-duality of the initial model is less surprising. It is possible to omit
the adjunction $F \dashv G$ from the structure of the model, and instead
simply specify the object $\CI$. The adjunction is then recovered using
the requirement that $\mathcal{C}$ have \emph{copowers}
(a concept from  enriched category theory), which is part of the
assumed structure of a model. The operation of taking the dual of a pointed
model, with point $\CR$, then has a simple description: instead of redefining the
adjunction, the r\^{o}les of the objects $\CI$ and $\CR$ are simply swapped in the structure.

Detailed definitions and proofs of 
all the semantic facts referred to above in this section will appear in
a paper devoted entirely to the category-theoretic models of EEC~\cite{EMSc}.
Unfortunately, although  the high-level ideas are straightforward, 
considerable technicalities arise in getting the details correct. 
The reader who wishes to see a slightly fuller treatment
than the outline given above, but not all details, is referred to the conference
version of the present paper~\cite{EMS:fossacs}.

To finish, we return to syntax. The alternative formulation of models, referred to
above, has a syntactic counterpart. Since $\CI$ is included as a primitive
computation type in our formulation of EEC, it would be possible
to omit, from the syntax of EEC, 
both the  type constructor $\Cbang{\VA}$ and the inclusion
of computation types amongst value types. The former can
be defined as $\Ccopower{\VA}{\CI}$. And the  value type corresponding to
a computation type $\CA$ can be recovered as  $\CI \lpop \CA$. (Thus Levy's
$U$ constructor~\cite{Levy:book} is rendered visible.) Using this 
restricted syntax, the involution property of Theorem~\ref{theorem:involution}
has a simplified form. There is no longer any need for the isomorphisms
$\Viso{\VA}$ and $\Ciso{\CA}$, since one obtains identities
$\CpsVVT{\VA} = \VA$ and $\CpsCCT{\CA} = \CA$. 
The very mild drawback of this formulation is that it 
requires the slightly more complex definition of
$\cbv{(\sigma \Lfun \tau)}  = \cbv{\sigma} \Vfun (\CI \lpop {\Ccopower{\cbv{\tau}\!}{\CI}})$
in Figure~\ref{figure:cbv:cbn}. Or alternatively, one could take
$\cbv{(\sigma \Lfun \tau)}  = {\Ccopower{\cbv{\sigma}\!}{\CI}} \lpop {\Ccopower{\cbv{\tau}\!}{\CI}}$,
which would fit in with redefining
$\cbvLincps{(\sigma \Lfun \tau)}  = 
(\cbvLincps{\tau} \Cfun \CR) \lpop (\cbvLincps{\sigma} \Cfun  \CR)$, as
discussed in Section~\ref{sec:lin:cps},
leaving the 
value-type-function-space constructor, $\Vfun$, superfluous to the translations.

However, for the present paper, we have preferred to retain $\Cbang{(\cdot)}$
as a primitive type construct, due to the basic r\^{o}le it plays in
related type systems: as $T$ in Moggi's computational metalanguage~\cite{Moggi:91},
as $F$ in Levy's CBPV~\cite{Levy:book}, and as the exponential in
linear logic~\cite{Girard:87}. For one thing, our choice of primitives
has allowed us to give the various translations of 
Sections~\ref{section:cbv:cbn} and~\ref{sec:lin:cps}
just as they appear in the literature
\cite{Moggi:91,Filinski:phd,Levy:book,Hasegawa:Flops:02,Hasegawa:Flops:04},
modulo the change  to EEC notation.
We also comment that it is perhaps the standard focus on 
$\Cbang{(\cdot)}$ (or $T$ or $F$) as the key construct in effect languages
that makes the involution property of the self-translation translation 
come as a surprise when first encountered. For, amongst computation types,
the type $\Cbang{\VA}$ has the most interesting translation --- the only one which
is not part of a dual pair. It is for this reason that the proofs of
Section~\ref{section:canonical} mainly focus on constructs associated with types of the form
$\Cbang{\VA}$ as providing  the interesting cases.

In the present paper, we have  investigated the enriched effect calculus  as a metalanguage for 
formalising one possible interaction between linearity and CPS translations.
It is the belief of the authors that EEC will prove
a useful language for modelling other ways in which linearity
and effects combine. Some potential examples of such interactions
are briefly discussed in the main paper introducing EEC~\cite{EMSb}. 
It would be interesting to see further convincing examples worked out
in detail.

\section*{Acknowledgements} 
We thank Masahito Hasegawa, Paul Levy and 
the anonymous referees for helpful suggestions.

\bibliographystyle{alpha}

\end{document}